\documentclass[journal]{IEEEtran}
\usepackage{graphicx,amssymb,amsmath}
\usepackage{multicol}
\usepackage{CJK}
\usepackage[noadjust]{cite}
\usepackage{setspace}
\usepackage{stfloats}
\usepackage{midfloat}
\usepackage{flushend,cuted}
\usepackage{bm}
\usepackage{multirow}
\usepackage{array}
\usepackage{stfloats}
\usepackage{cite}
\usepackage{algorithm}
\usepackage{algpseudocode}
\usepackage{psfrag}
\usepackage[caption=false]{subfig}
\usepackage{color}

\IEEEoverridecommandlockouts

\ifCLASSINFOpdf
\else
\fi
\newcommand{\tr}{\textmd{tr}}
\newcommand{\E}{\textmd{E}}
\newcommand{\Var}{\textmd{Var}}
\newcommand{\diag}{\textmd{diag}}
\newcommand{\vect}{\textmd{vec}}
\newcounter{MYtempeqncnt}

\allowdisplaybreaks
\begin{document}
\title{Optimal Design of Energy and Spectral Efficiency Tradeoff in One-Bit Massive MIMO Systems}


\author{Yongzhi Li,
        Cheng Tao,~\IEEEmembership{Member, IEEE,}
        Amine Mezghani,
        A. Lee Swindlehurst,~\IEEEmembership{Fellow,~IEEE}\\
        Gonzalo Seco-Granados,~\IEEEmembership{Senior Member, IEEE,}
        ~and~Liu Liu
\thanks{Y. Li, C. Tao, and L. Liu are with the Institute of Broadband
Wireless Mobile Communications, School of Electronic and Information
Engineering, Beijing Jiaotong University, Beijing, 100044, China (email:
liyongzhi@bjtu.edu.cn; chtao@bjtu.edu.cn; liuliu@bjtu.edu.cn).}
\thanks{G. Seco-Granados is with the Telecommunications and Systems
Engineering Department, Universitat Aut${\rm{\grave{o}}}$noma de Barcelona, Barcelona
08193, Spain (e-mail: gonzalo.seco@uab.es).}
\thanks{A. Mezghani and A. L. Swindlehurst are with the Center for Pervasive Communications and
Computing, University of California, Irvine, CA 92697 USA (e-mail: amezghan@uci.edu; swindle@uci.edu). A. L. Swindlehurst is also a Hans Fischer Senior Fellow of the Institute for Advanced Study at the Technical University of Munich.}
\thanks{The research was supported in part by the National 863 Project Granted No. 2014AA01A706, Beijing Natural Science Foundation project Grant No. 2016023, Fundamental Research Funds for the Central Universities under grant 2015JBM011, the NSFC project under grant No. 61471027, the Research Fund of National Mobile Communications Research Laboratory, Southeast University No. 2014D05, and Beijing Natural Science Foundation project under grant No.4152043. A.~Swindlehurst was supported by the National Science Foundation under Grant ECCS-1547155, and by the Technische Universit\"at M\"unchen Institute for Advanced Study, funded by the German Excellence Initiative and the European Union Seventh Framework Programme under grant agreement No. 291763, and by the European Union under the Marie Curie COFUND Program.}
}

\maketitle

\begin{abstract}
This paper considers a single-cell massive multiple-input multiple-output (MIMO) system {equipped with a base station (BS) that uses one-bit quantization} and investigates the energy efficiency (EE) and spectral efficiency (SE) trade-off. We first propose a new precoding scheme and downlink power allocation {strategy that results in uplink-downlink SINR duality for} one-bit MIMO systems. Taking into account the effect of the imperfect channel state information, we obtain approximate closed-form expressions for the uplink {and downlink} achievable rates {under duality} with maximum ratio combining/matched-filter and zero-forcing processing. 
{We then focus on joint optimization of the competing SE and EE objectives over the number of users, pilot training duration and operating power, using the weighted product method to obtain the EE/SE Pareto boundary.}
Numerical results are presented to verify our analytical results and demonstrate the fundamental tradeoff between EE and SE for different parameter settings.
\end{abstract}


%
\IEEEpeerreviewmaketitle

\section{Introduction}
In the past decade, there has been {considerable work focused on improving wireless} system spectral efficiency (SE) and throughput {triggered by} spectrum scarcity and the demand for higher {rates for} multimedia applications. Massive multiple-input multiple-output (MIMO) technology {is considered to be an important component of 5th} generation (5G) wireless communication systems, {and has been shown to potentially achieve increases in} SE by orders of magnitude over contemporary systems. The main idea of massive MIMO is based on equipping {the} base stations (BSs) with many antenna elements, {providing} unprecedented spatial degrees of freedom for simultaneously serving multiple user terminals on the same time-frequency channel \cite{marzetta2010noncooperative, lu2014overview,larsson2013massive,rusek2013scaling}.

However, with a large number of antenna elements deployed at the BS, system cost and power consumption will be excessive if each antenna element and corresponding {\mbox{radio}} frequency (RF) chain is equipped with high-resolution and power-hungry analog-to-digital converter/digital-to-analog converters (ADC/DACs). In addition, as huge bandwidths and correspondingly high sampling rates will be required in next generation wireless systems, high-speed ADCs are either unavailable or too costly for practical implementation \cite{murmann2016adc}. Therefore, the use of {low-resolution ADC/DACs, especially those that operate with only one bit}, has been suggested for massive MIMO systems \cite{jiayi2016on,chiara2014massive,juncil2015near,yongzhi2016channel,mollen2016performance}.
{One-bit ADC/DACs significantly reduce power consumption and cost since they consist of a simple comparator and do not require automatic gain control or linear amplifiers, and hence greatly simplify the analog RF front end.}
It has been shown that MIMO capacity is not severely reduced by the coarse quantization at low signal-to-noise ratios (SNRs) \cite{mezghani2008analysis}; the power penalty due to one-bit quantization is approximately equal to only $\pi/2$ (1.96dB) in the low SNR region \cite{nossek2006capacity}. In addition, \cite{yongzhi2016channel} and \cite{mollen2016performance} showed that, compared with the conventional MIMO systems with perfect hardware implementation, the loss of SE in one-bit MIMO systems can be compensated {for by} with deploying around 2.2-2.5 times more antennas at the BS. Therefore the use of one-bit ADC/DACs can make massive MIMO systems more viable in practice.

{While SE is the most common performance metric for designing wireless systems \cite{emil2016massive,ngo2014massive}, recent efforts to promote energy efficient or "green" communications have recently attracted considerable research interest \cite{han2011green,chen2011fundamental}.}
There has been a large amount of work devoted to maximizing the EE for massive MIMO in various scenarios, including single cell and multi-cell deployments \cite{ngo2013energy,emil2015optimal}. {SE and EE are typically competing objectives; increasing one usually leads to degradations in the other. Since both are important factors in wireless {systems}, it is important to consider both when optimizing the system design. Multiple objective optimization approaches are useful for obtaining reasonable operating points on the Pareto performance boundary that balance the two criteria \cite{marler2004survey,branke2008multiobjective,bjornson2014multiobjective}.}

There has been limited prior work on the EE-SE tradeoff in various wireless communication systems \cite{tang2014resource,he2013energy,zhang2016spectral,hao2015energy}. {The} authors of \cite{tang2014resource} proposed a new paradigm for the EE-SE tradeoff in OFDMA cellular networks and showed that a significant amount of bandwidth can be saved with a slight increase in energy consumption by using the proposed paradigm. In \cite{he2013energy}, the relationship between EE and SE for  downlink multiuser distributed antennas systems (DAS) is investigated, and an algorithm to allocate the available power to balance EE and SE is also proposed. {A full duplex relay system was considered in [23], and several power scaling schemes were proposed that achieve a good trade-off between SE and EE.}
{The work of} \cite{hao2015energy} investigated the SE-EE tradeoff {using} multi-objective optimization subject to a maximum total transmit power constraint for downlink massive MIMO systems for {several} different parameter settings.

In this paper, we consider a single-cell massive MIMO {system} with a BS equipped with one-bit {ADC/DACs} and {we simultaneously} address the EE-SE tradeoff {for both} uplink and downlink. Our goal is to provide insight on how the number of active terminals, the pilot training length and the system operating power affect the EE and SE for different linear processing schemes. Our specific contributions are summarized below.
\begin{itemize}
  \item Since uplink-downlink duality is helpful when jointly designing the precoder and receiver, we prove the
      uplink-downlink SINR duality for one-bit MIMO systems {operating at low SNR}. On the basis of this duality, we propose a new precoding scheme and downlink power allocation {strategy that achieve the same SINR in both the uplink and downlink.} {In addition, we show that by using our proposed power allocation strategy, the {required} range of power amplifier gains is small, especially for a large number of transmit antennas. }
  \item We quantify the theoretical rate achievable in both the {\mbox{uplink}} and downlink using MRC/MF and ZF processing based on { linear minimum mean-squared error (LMMSE)} channel estimation. Closed-form expressions {for} the uplink and downlink achievable rates with {MRC/MF and ZF} are obtained as well, which enable us to {simultaneously examine} the uplink and downlink performance to evaluate the system SE and EE.
  \item {Using a standard} multiple objective optimization {approach}, we investigate the optimal number of active terminals, pilot training length and system operating power {that achieve a good tradeoff between SE and EE, and we show the benefit of doing so via several numerical examples.} {We further} show that the power penalty in one-bit systems can be compensated {for} by deploying 2-2.5 times more antennas for MRC/MF processing, while for ZF processing, more and more antennas are required as the SNR increases.
\end{itemize}

The rest of this paper is organized as follows. In the next section, we present the assumed system architecture and signal model for both the uplink and downlink. In Section III, we investigate channel estimation and data transmission for one-bit MIMO systems using the Bussgang decomposition. The uplink-downlink SINR duality in one-bit MIMO systems is proved in Section IV. In Section V, we provide approximate expressions for the uplink and downlink achievable rate. We focus on the energy and spectral efficiency trade-off in Section VI, and investigate the {effect of} optimal resource allocation {in improving both the SE and EE.} Simulation results are presented in Section VI and we conclude the paper in Section VII.

\emph{Notation}: The following notation is used throughout the paper. Bold uppercase (lowercase) letters denote matrices (vectors); $(.)^*$, $(.)^T$, and $(.)^H$ denote complex conjugate, transpose, and Hermitian transpose operations, respectively; $||.||$ represents the 2-norm of a vector; $\tr(.)$ represents the trace of a matrix; $\diag\{\mathbf{X}\}$ denotes a diagonal matrix containing only the diagonal entries of $\mathbf{X}$; $\otimes$ represents the Kronecker product; $\left[\mathbf{X}\right]_{ij}$ denotes the $(i,j)$th entry of $\mathbf{X}$; ${\bf{x}} \sim \mathcal{CN}\left( {{\bf{a}},{\bf{B}}} \right)$ indicates that $\bf{x}$ is a complex Gaussian vector with mean $\bf{a}$ and covariance matrix $\bf{B}$; $\E\{.\}$ and $\Var\{.\}$ denote the expected value and variance of a random variable, respectively.

\section{System Model}

\begin{figure}
  \centering
  \includegraphics[width=8cm]{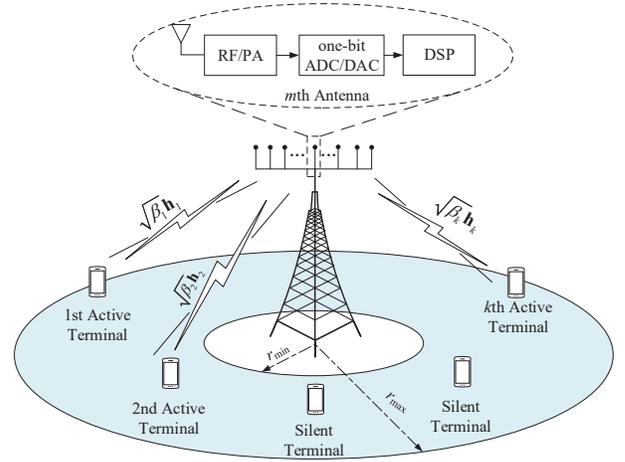}\\
  \caption{System architecture of one-bit massive MIMO system.}\label{1bit}
  \vspace{-0.3cm}
\end{figure}

As depicted in Fig.~\ref{1bit},
we consider a single-cell massive MIMO system, {where each antenna of the base station (BS) is assumed to be equipped with {one-bit ADC/DACs.} For the uplink, the one-bit ADCs convert the real and imaginary part of the received signal from the analog domain, while for the downlink transmission, the one-bit DACs convert the precoded data symbols fed to the analog domain.} We assume the BS is deployed with a large array of $M$  antennas and communicates with $K\gg 1$ out of $K_{\rm max}$ single-antenna terminals at a time, where we refer {to} the $K$ terminals as {\it active terminals} and {to} the remaining $K_{\rm max} - K$ as {\it silent terminals}. The $K$ active terminals are randomly selected in a round-robin fashion, such that the subset of active terminals changes over time, and hence the index of the terminal $k\in\{1,...,K\}$ {corresponds} to different terminals at different {times}. The geographical position of the $k$th terminal is {assumed to be} a random variable {drawn from} a {circular uniform distribution with maximum radius $r_{\rm max}$ and minimum radius $r_{\rm min}$}.

Let $\mathbf{g}_k\in\mathbb{C}^{M\times 1}$ denote the channel vector between the $k$th active terminal and the BS, which {includes the effects of} independent fast fading, geometric attenuation and log-normal shadow fading:
\begin{equation}\label{channel_model}
\mathbf{g}_k = \sqrt{\beta_k} \mathbf{h}_k,
\end{equation}
where $\mathbf{h}_k\sim\mathcal{CN}(\mathbf{0},\mathbf{I})$ represents the fast fading channel of the $k$th active terminal and is assumed to be Gaussian distributed with zero mean and unit variance, and $\beta_k$ is the large-scale fading coefficient that models both the geometric attenuation and shadow fading.
The large scale fading coefficient of the $k$th active terminal is modeled as
\begin{equation}\label{large_scale_model}
\beta_k = \frac{\bar{d}}{(d_k/r_{\min})^\kappa},
\end{equation}
where $\bar{d}$ is a non-logarithmic shadowing value, $\kappa$ is the path-loss exponent and $d_k$ represents the distance between the $k$th user and the BS.  Due to the slow-varying nature of the long term channel statistics, we assume the large scale fading coefficients are known at the BS \cite{marzetta2010noncooperative,ngo2013energy}.

We assume {that} the one-bit massive MIMO system operates in time-division duplex (TDD) mode and the channels of the $K$ active terminals {undergo} block fading {in which} the channel {state remains} constant over a coherence interval of length $T$ symbols. {Assuming reciprocal} uplink and downlink channels, the BS can process both the uplink and downlink signals using the uplink channel measurement. 

\subsection{Uplink Signal Model}
{During the uplink phase}, we assume the $K$ active terminals simultaneously transmit independent data symbols to the BS. Thus the uplink received signal at the BS is
\begin{equation}\label{uplink_signal_model}
{\mathbf{y}}_{\rm u} = \sum_{k=1}^{K} \sqrt{p_{k}} \mathbf{g}_{k} s_k + {\mathbf{n}_{\rm u}},
\end{equation}
where the subscript `${\rm u}$' stands for `uplink', $p_k$ and $s_k$ {are respectively} the transmit power and transmit data symbol of the $k$th active terminal. In addition we assume $\E\{|s_k|^2\} = 1$ and $\E\{s_i s_k^H\} = 0 $ for $i \neq k$. The additive noise is assumed to be Gaussian: $\mathbf{n}_{\rm u}\sim\mathcal{CN}(0,\mathbf{I})\in\mathbb{C}^{M\times 1}$. {The total uplink} transmit power is then given by $P_{\rm total}^{\rm (UL)} = \sum_{k=1}^K p_k$.

In most existing work, fixed and equal uplink transmit power is assumed for each active terminal. This may result in a large rate disparity between the nearest and furthest active terminals {due to path loss}. However, we {assume a} statistically-aware power control strategy {in which} the data symbols from the $k$th terminal have transmit power $ p_k = \beta_k^{-1}\rho_{\rm u}$, where $\rho_{\rm u}$ is a system design parameter we refer to as the system {\it operating power} \cite{emil2016massive}. The quantized signal obtained after the one-bit ADCs is represented as
\begin{align}\label{uplink_signal_model}
\mathbf{r}_{\rm u} &= \mathcal{Q}(\mathbf{y}_{\rm u}) = \mathcal{Q}\left(\sum_{k=1}^{K} \sqrt{p_k} \mathbf{g}_{k} s_k + {\mathbf{n}_{\rm u}}\right) \nonumber \\
& = \mathcal{Q}\left(\sum_{k=1}^{K} \mathbf{g}^{\rm eff}_{k} s_k + {\mathbf{n}_{\rm u}}\right),
\end{align}
where $\mathbf{g}_k^{\rm eff} = \sqrt{p_k}\mathbf{g}_k = \sqrt{\rho_{\rm u}}\mathbf{h}_k$ represents the effective power-controlled channel of the $k$th active terminal. Since the large scale fading coefficients can be compensated for by adjusting the transmit power, we model the effective channel as $\mathbf{g}_k^{\rm eff}\sim\mathcal{CN}(0,\rho_{\rm u}\mathbf{I})$, which implies that a uniform terminal experience can be guaranteed since each active terminal has the same channel statistics. {The operator} $\mathcal{Q}(.)$ represents the one-bit quantization, which is applied separately to the real and imaginary parts as $\mathcal{Q}(.) = \frac{1}{\sqrt{2}} \left(\textrm{sign}\left(\Re\left(.\right)\right)+ j\textrm{sign}\left(\Im\left(.\right)\right)\right)$. Thus, the output set of the one-bit quantization is equivalent to the quadrature phase-shift keying (QPSK) constellation points ${\Omega} = \frac{1}{\sqrt{2}}\{1+j, 1-j, -1+j, -1-j\}$.

\subsection{Downlink Signal Model}
For the downlink, we assume
{that the data symbol $x_k$ for the $k$-th active terminal is multiplied by a precoder $\mathbf{t}_k$, and then the real and imaginary parts are quantized by a one-bit DAC:}
\begin{equation}\label{r_d}
\mathbf{r}_{\rm d} = \mathcal{Q}\left(\mathbf{y}_{\rm d}\right)= \mathcal{Q}\left({\sum_{k=1}^{K} \mathbf{t}_k x_k}\right),
\end{equation}
where the subscript `${\rm d}$' stands for `downlink', {the data symbols} $x_k$ are zero mean with unit variance and satisfy $\E\{|x_k|^2\}=1$ and $\E\{x_i x_k^H\} = 0$ for $i\neq k$. Again, each element of $\mathbf{r}_{\rm d}$ lies in the set $\Omega$ due to the one-bit quantization. As we assume that the system operates in TDD mode, 
the received signal at the $k$th active terminal is given by
\begin{align}\label{downlink_signal_model}
u_{{\rm d},k} &= \mathbf{g}_k^T \mathbf{Q}\mathbf{r}_{\rm d} + n_{{\rm d},k} \nonumber \\
&= \mathbf{g}_k^T \mathbf{Q}\mathcal{Q}\left({\sum_{k=1}^{K} \mathbf{t}_k x_k}\right) + n_{{\rm d},k},
\end{align}
where $n_{{\rm d},k}\sim\mathcal{CN}(0,1)$ is additive white Gaussian noise at the $k$th active terminal, $\mathbf{Q}$ is a diagonal matrix that represents the power {allocated to} each BS antenna, and hence the total transmit power for the downlink is $P_{\rm total}^{\rm (DL)} = \tr\left(\mathbf{Q}\mathbf{Q}^H\right)$.

Note that contrary to conventional MIMO systems where the transmit power allocation can be considered jointly with the precoding vector, {the power allocated to} each BS antenna {in a one-bit MIMO system} should be considered separately since the power of each element of the output signal $\mathbf{r}_{\rm d}$ is normalized due to the one-bit quantization.  We later show how to design the precoding vector $\mathbf{t}_k$ and the transmit power matrix $\mathbf{Q}$ to achieve the same achievable rate performance in the downlink as in the uplink.

\section{Channel Estimation and Data Transmission for One-Bit MIMO Systems}
We assume each {coherence} interval $T$ is divided into three parts: the first $\tau = \tau_0 K$ symbols reserved for pilot training, and then the remaining $T-\tau $ symbols are split {between} the uplink and downlink data transmission phases.
{The factor $\tau_0$ represents the number of pilots relative to the number of active terminals. In conventional MIMO systems, the value that maximizes SE is $\tau_0=1$, while for one-bit systems SE is in general maximized for $\tau_0 > 1$ \cite{yongzhi2016channel}.}
We denote $\gamma\in(0,1)$ as the {fraction of the} coherence interval allocated for uplink transmission and $1-\gamma$ the fraction for downlink transmission.

\subsection{Channel Estimation}
{We assume the CSI is estimated based on pilot sequences of length $\tau$ symbols:}
\begin{equation}\label{training_received_signal}
\mathbf{Y}_{\rm t} = \sum_{k=1}^K \sqrt{p_k}\mathbf{g}_k\bm{\phi}_k^T +\mathbf{N}_{\rm t} = \sum_{k=1}^K \mathbf{g}_k^{\rm eff}\bm{\phi}_k^T +\mathbf{N}_{\rm t},
\end{equation}
where the subscript `${\rm t}$' stands for `training', $\mathbf{Y}_{\rm t}\in\mathbb{C}^{M\times\tau}$ is the received training signal matrix, $p_k$ and $\bm{\phi}_k\in\mathbb{C}^{\tau\times 1}$ are the pilot transmit power and the pilot sequence of the $k$th user, respectively.  In this paper, {all pilot sequences are assumed to be mutual orthogonal}, i.e., $\bm{\phi}_k^T\bm{\phi}_k^* = \tau$ and $\bm{\phi}_i^T\bm{\phi}_k^* = 0$ for $i\neq k$.

To match the matrix form of \eqref{training_received_signal} to the vector form of \eqref{uplink_signal_model}, we vectorize the received signal as
\begin{equation}\label{vec_training_received_signal}
{\mathbf{y}_{\rm t} = \vect(\mathbf{Y}_{\rm t}) =} \sum_{k=1}^K\bar{\bm{\phi}}_k{\mathbf{g}}_{k}^{\rm eff} + {\mathbf{n}}_{\rm t},
\end{equation}
where $\bar{\bm{\phi}}_k = \left(\bm{\phi}_k \otimes \mathbf{I}_{M}\right)$ and ${\mathbf{n}}_{\rm t} = \vect(\mathbf{N}_{\rm t})$. After one-bit ADCs, the quantized signal is expressed as
\begin{align}\label{training_quantized_signal}
  \mathbf{r}_{\rm t} = \mathcal{Q}(\mathbf{y}_{\rm t}) = \mathcal{Q}\left(\sum_{k=1}^K\bar{\bm{\phi}}_k{\mathbf{g}}_k^{\rm eff} + {\mathbf{n}}_{\rm t}\right),
\end{align}
where each element of $\mathbf{r}_{\rm t}$ takes values from the set $\Omega$.
{We would like to remark that the output of \eqref{training_quantized_signal}, $\mathbf{r}_{\rm t}$, is independent of any real-valued scaling factor applied to $\mathbf{y}_{\rm t}$}.
Using the Bussgang decomposition \cite{bussgang1952yq}, we can reformulate the nonlinear quantization using a statistically equivalent linear operator that will simplify the channel estimator and the resulting analysis:
\begin{equation}\label{r_t_Bussgang}
  \mathbf{r}_{\rm t} = \mathcal{Q}(\mathbf{y}_{\rm t}) = \mathbf{A}_{\rm t} \mathbf{y}_{\rm t} + \bm{\eta}_{\rm t} = \sum_{k=1}^K{\tilde{\bm{\phi}}}_k \mathbf{g}_k^{\rm eff} + \tilde{\mathbf{n}}_{\rm t},
\end{equation}
where {${\bm{\eta}}_{\rm t}$ is the quantization noise,} ${\tilde{\bm{\phi}}}_k = \mathbf{A}_{\rm t} {\bar{\bm{\phi}}}_k$, $\tilde{\mathbf{n}}_{\rm t} = \mathbf{A}_{\rm t} \mathbf{n}_{\rm t} + \bm{\eta}_{\rm t}$ is the total effective noise, and $\mathbf{A}_{\rm t}$ is chosen to make $\bm{\eta}_{\rm t}$ uncorrelated with $\mathbf{y}_{\rm t}$ \cite{bussgang1952yq}, or equivalently, to minimize the power of the equivalent quantization noise. {If the received signal $\mathbf{y}_{\rm t}$ is Gaussian, then $\mathbf{A}_{\rm t}$ {is} given by} \cite{yongzhi2016channel}
\begin{align}\label{A_t}
\mathbf{A}_{\rm t} &= \sqrt{\frac{2}{\pi}}\diag\left(\mathbf{C}_{\mathbf{y}_{\rm t}}\right)^{-\frac{1}{2}} \nonumber \\
& = \sqrt{\frac{2}{\pi}}\diag\left(\sum_{k=1}^K\rho_{\rm u}(\bm{\phi}_k\bm{\phi}_k^H\otimes \mathbf{I}_M) + \mathbf{I}_{M\tau}\right)^{-\frac{1}{2}},
\end{align}
where $\mathbf{C}_{\mathbf{y}_{\rm t}}$ denotes the covariance matrix of $\mathbf{y}_{\rm t}$.

We can see from \eqref{A_t} that $\mathbf{A}_{\rm t}$ is related to the diagonal terms of $\mathbf{C}_{\mathbf{y}_{\rm t}}$ and therefore, to the pilot sequences. {In order to obtain a simple expression for $\mathbf{A}_{\rm t}$,
we consider pilot sequences composed of submatrices of the discrete Fourier transform (DFT) operator \cite{biguesh2004downlink}.} The benefits of using DFT pilot sequences are: i) all elements of the pilot sequences have the same magnitude, which simplifies peak transmit power constraints, and ii) the diagonal terms of $\bm{\phi}_k\bm{\phi}_k^H$ are always equal to 1, which results in {the following} simple expression for $\mathbf{A}_{\rm t}$:
\begin{equation}\label{A_t2}
\mathbf{A}_{\rm t} = \sqrt{\frac{2}{\pi}\frac{1}{K\rho_{\rm u} + 1}}\mathbf{I} \triangleq \alpha\mathbf{I}.
\end{equation}

Since the one-bit quantization always results in a signal $\mathbf{r}_{\rm t}$ with unit variance, the individual transmit power $p_k$ of each user is {difficult to estimate} \cite{bar2002doa}. Therefore we estimate the effective channels $\mathbf{g}_{k}^{\rm eff}$ including the uplink power-control instead of the true channel $\mathbf{g}_k$. In this way, we can estimate the channel without taking the transmit power $p_k$ into account. According to \cite{kay1993fundamentals} and the fact that $\bm{\eta}_p$ is uncorrelated with the channel ${\mathbf{g}_k^{\rm eff}}$ (see \cite{yongzhi2016channel} for a detailed proof), the LMMSE channel estimate of ${\mathbf{g}_k^{\rm eff}}$ can be expressed as
\begin{equation}\label{channel_estimate}
\hat{\mathbf{g}}_k^{\rm eff} = \rho_{\rm u}\tilde{\bm{\phi}}_k^H\mathbf{C}_{\mathbf{r}_{\rm t}}^{-1}\mathbf{r}_{\rm t},
\end{equation}
where $\mathbf{C}_{\mathbf{r}_{\rm t}}$ is the auto-correlation matrix of $\mathbf{r}_{\rm t}$. It has been shown in \cite{jacovitti1994estimation} that for one-bit ADCs, the arcsine law can be used to obtain $\mathbf{C}_{\mathbf{r}_{\rm t}}$:
\begin{align}\label{C_rr}
\mathbf{C}_{\mathbf{r}_{\rm t}} =& \frac{2}{\pi} \left(\arcsin\left(\bm{\Sigma}^{-\frac{1}{2}}_{\mathbf{y}_{\rm t}} \Re\left({\mathbf{C}_{\mathbf{y}_{\rm t}}}\right) \bm{\Sigma}^{-\frac{1}{2}}_{\mathbf{y}_{\rm t}} \right) \right. \nonumber \\
& \left. + j \arcsin\left(\bm{\Sigma}^{-\frac{1}{2}}_{\mathbf{y}_{\rm t}} \Im\left({\mathbf{C}_{\mathbf{y}_{\rm t}}}\right) \bm{\Sigma}^{-\frac{1}{2}}_{\mathbf{y}_{\rm t}} \right)  \right),
\end{align}
where we define $\bm{\Sigma}_{\mathbf{y}_{\rm t}} = \diag\left(\mathbf{C}_{\mathbf{y}_{\rm t}}\right)$. The covariance matrix of the estimation error  $\bm{\varepsilon}_k^{\rm eff} = \mathbf{g}_k^{\rm eff} - \hat{\mathbf{g}}_k^{\rm eff}$ is then given by
\begin{align}\label{channel_estimate_error_cov}
\mathbf{C}_{\bm{\varepsilon}_k^{\rm eff}} &= \E\left\{\left({\mathbf{g}}_k^{\rm eff} - \hat{\mathbf{g}}_k^{\rm eff}\right)\left({\mathbf{g}}_k^{\rm eff} - \hat{\mathbf{g}}_k^{\rm eff}\right)^H\right\} \nonumber \\
& = \rho_{\rm u}\mathbf{I}_M - \rho_{\rm u}^2\tilde{\bm{\phi}}_k^H\mathbf{C}_{\mathbf{r}_{\rm t}}^{-1}\tilde{\bm{\phi}}_k
\end{align}
and the mean-squared error (MSE) is ${\rm MSE}_k = \tr(\mathbf{C}_{\bm{\varepsilon}_k^{\rm eff}})$.

{Note that $\mathbf{C}_{\mathbf{y}_{\rm t}}$ is a circulant matrix, and hence $\mathbf{C}_{\mathbf{r}_{\rm t}}$ is also circulant since the arcsine operation is performed element-wise. Note also that the inverse of a circulant matrix is also circulant, and that circulant matrices can be diagonalized by a DFT matrix.}
Therefore, another advantage of using the DFT matrix as pilot sequences is that, by using the channel estimator in \eqref{channel_estimate}, the elements of the channel estimate are mutual uncorrelated. Thus, the covariance matrix of the estimation error in \eqref{channel_estimate_error_cov} is diagonal.


\subsection{Data Transmission}
During the data transmission phase, the BS considers the channel estimates as the true channel and employs
{a linear receiver to decode the uplink signals from the $K$ active terminals, and a linear precoder to broadcast the downlink signal to the $K$ terminals. }
\subsubsection{Linear Receiver}
During the uplink, the $K$ users simultaneously transmit their data symbols
to the BS. We again employ the Bussgang decomposition and reformulate the nonlinear quantizer $\mathcal{Q}(.)$ as a linear function, and the quantized signal at the BS can be expressed as
\begin{align}\label{r_u_Bussgang}
\mathbf{r}_{\rm u} &= \mathcal{Q}(\mathbf{y}_{\rm u})=\mathcal{Q}\left(\sum_{k=1}^K\sqrt{p_k}\mathbf{g}_k s_k +  \mathbf{n}_{\rm u}\right)  \nonumber \\
&= \mathbf{A}_{\rm u}\sum_{k=1}^K \mathbf{g}_k^{\rm eff} s_k + \mathbf{A}_{\rm u}\mathbf{n}_{\rm u} + \bm{\eta}_{\rm u},
\end{align}
where the same definitions as in the previous sections apply, but replacing the subscript `${\rm t}$' with `${\rm u}$'. Again, according to the Bussgang decomposition, the matrix $\mathbf{A}_{\rm u}$ for a Gaussian input is given by
\begin{align}\label{A_u}
\mathbf{A}_{\rm u} &= \sqrt{\frac{2}{\pi}}\diag\left(\mathbf{C}_{\mathbf{y}_{\rm u}}\right)^{-\frac{1}{2}} \nonumber \\
& = \sqrt{\frac{2}{\pi}}\diag\left(\sum_{k=1}^K\mathbf{g}_k^{\rm eff}(\mathbf{g}_k^{\rm eff})^H + \mathbf{I}_{M}\right)^{-\frac{1}{2}} \nonumber \\
& \cong \sqrt{\frac{2}{\pi}}\sqrt{\frac{1}{K \rho_{\rm u} + 1}}\mathbf{I}_M = \alpha\mathbf{I}_M,
\end{align}
where the approximation is reasonable since we assume $K \gg 1$ and {i.i.d.} effective channel coefficients.

Note that {unlike} \eqref{A_t}, where the pilot sequences are fixed and the expectation operation is taken over the channel vectors, the expectation in \eqref{A_u} is taken over the data symbols for each individual channel realization. {As such}, the effect that the quantization noise sometimes appears to coherently combine with the data symbols {as observed in \cite{mollen2016performance}} can be avoided.

With a linear receiver, the quantized signal $\mathbf{r}_{\rm u}$ is separated into $K$ streams {via multiplication by the} matrix $\mathbf{W}^T$ as follows:
\begin{align}\label{Linear_Receiver}
{\bf{\hat s}} &= \mathbf{W}^T\mathbf{r}_{\rm u}\nonumber\\
&= \mathbf{W}^T\mathbf{A}_{\rm u}\sum_{k=1}^K {\mathbf{g}}_k^{\rm eff} s_k + \mathbf{W}^T\mathbf{A}_{\rm u}\mathbf{n}_{\rm u} + \mathbf{W}^T\bm{\eta}_{\rm u}.
\end{align}
Then the $k$th stream is used to decode the signal transmitted from the $k$th active terminal, which can be expressed as
\begin{align}\label{hat_s_k}
  {\hat s_k} =& { {\mathbf{{w}}}_k^T{{\bf{A}}_{\rm u}}{{\bf{g}}^{\rm eff}_k}{s_k}}+  {{\mathbf{{w}}}_k^T \sum\nolimits_{i \ne k}^K {{\bf{A}}_{\rm u}}{{\bf{g}}_i^{\rm eff}}{s_i}} \nonumber\\
  & + {{\mathbf{{w}}}_k^T{{\bf{A}}_{\rm u}}{\bf{n}}_{\rm u}} + {{\mathbf{{w}}}_k^T{\bm{\eta}}_{\rm u}},
\end{align}
where $\mathbf{w}_k$ is the $k$th column of $\mathbf{W}$. The last three terms in \eqref{hat_s_k} respectively correspond to user interference, AWGN noise and quantization noise.

\subsubsection{Linear Precoding}
{Similar to the uplink, in the downlink we can}
reformulate the precoded signal model of \eqref{r_d} using the Bussgang decomposition:
\begin{align}\label{r_d_Bussgang}
\mathbf{r}_{\rm d} &= \mathcal{Q}\left({\sum_{k=1}^{K} \mathbf{t}_k x_k}\right) = \mathbf{A}_{\rm d}\sum_{k=1}^{K} \mathbf{t}_k x_k +\bm{\eta}_{\rm d},
\end{align}
where the same definitions as in the previous sections apply, but replacing the subscript `${\rm t}$` with `${\rm d}$`. The matrix $\mathbf{A}_{\rm d}$  for a Gaussian input is given by
\begin{align}\label{A_d}
\mathbf{A}_{\rm d} = \sqrt{\frac{2}{\pi}}\diag\left(\mathbf{C}_{\mathbf{y}_{\rm d}}\right)^{-\frac{1}{2}} = \sqrt{\frac{2}{\pi}}\diag\left(\sum_{k=1}^{K} \mathbf{t}_k\mathbf{t}_k^H\right)^{-\frac{1}{2}}.
\end{align}
Substituting \eqref{r_d_Bussgang} into \eqref{downlink_signal_model}, the received signal at the $k$th terminal can be expressed as
\begin{align}\label{u_d_k}
&u_{{\rm d},k}= \mathbf{g}_k^T \mathbf{Q}\mathcal{Q}\left({\sum_{k=1}^{K} \mathbf{t}_k x_k}\right) + n_{{\rm d},k} \nonumber \\
& = \mathbf{g}_k^T \mathbf{Q}\mathbf{A}_{\rm d}\mathbf{t}_k x_k + \mathbf{g}_k^T \mathbf{Q}\mathbf{A}_{\rm d}{\sum_{i\neq k}^{K} \mathbf{t}_i x_i} + \mathbf{g}_k^T \mathbf{Q}\bm{\eta}_{\rm d}+ n_{{\rm d},k}.
\end{align}
The last three terms in \eqref{u_d_k} respectively correspond to multiuser interference, quantization noise and AWGN noise.

\section{Uplink-Downlink SINR Duality in \\One-Bit MIMO Systems}
The receiver and precoder can be designed in a variety of ways to satisfy specific uplink and downlink SINR {requirements}. In general, it is more difficult to optimize {the downlink} precoder than {the uplink} receiver, and the uplink-downlink SINR duality theorem is a key tool which {substantially simplifies} the problem of joint precoder and receiver {design}.
The duality between the downlink (broadcast) and the uplink (multiple access) {channels in} unquantized MIMO systems has been proved in previous work \cite{vishwanath2003duality}. {As with conventional MIMO systems, establishing SINR duality for the one-bit case is useful because, for example, it allows one to transform the more difficult problem of downlink precoder design and power allocation to the easier dual uplink problem.}
{Thus, in the following, we establish the conditions for which}
the uplink and downlink duality holds in one-bit systems. As we will see later, this uplink-downlink duality allows us to derive an EE-SE tradeoff that is valid for both links.

According to the signal model of \eqref{hat_s_k} and \eqref{u_d_k}, the uplink and downlink SINR of the $k$th active user can be expressed as \eqref{SINR_UL} and \eqref{SINR_DL} shown on the top of next page,
\begin{figure*}[!t]
\normalsize
\setcounter{MYtempeqncnt}{\value{equation}}
\setcounter{equation}{22}
\begin{align}
\label{SINR_UL}{\rm SINR}_{k}^{\rm (UL)} &= \frac{p_k|{\mathbf{{w}}}_k^T{{\bf{A}}_{\rm u}}{{\bf{g}}_k}|^2}{\sum_{i \ne k}^K p_i|{{\mathbf{{w}}}_k^T {{\bf{A}}_{\rm u}}{{\bf{g}}_i}}|^2  + \|{\mathbf{{w}}}_k^T{{\bf{A}}_{\rm u}}\|_2^2 + {\mathbf{{w}}}_k^T\mathbf{C}_{{\bm{\eta}}_{\rm u}}{\mathbf{{w}}}_k^*}, \\
\label{SINR_DL}{\rm SINR}_{k}^{\rm (DL)} &=\frac{|\mathbf{g}_k^T \mathbf{Q}\mathbf{A}_{\rm d}\mathbf{t}_k|^2}{\sum_{i\neq k}^{K}|\mathbf{g}_k^T \mathbf{Q}\mathbf{A}_{\rm d} \mathbf{t}_i |^2 + \mathbf{g}_k^T \mathbf{Q}\mathbf{C}_{{\bm{\eta}}_{\rm d}}\mathbf{Q}^H\mathbf{g}_k^*+ 1}.
\end{align}
\setcounter{equation}{24}
\hrulefill
\vspace*{-0.3cm}
\end{figure*}
where $\mathbf{C}_{{\bm{\eta}}_{\rm p}}$  denotes the covariance matrix of the quantizer noise ${\bm{\eta}}_{\rm p }$
\begin{equation}\label{quantizer_Covariance_Matrix}
\mathbf{C}_{{\bm{\eta}}_{\rm p}}  = \mathbf{C}_{\mathbf{r}_{\rm p}} - \mathbf{A}_{\rm p}\mathbf{C}_{\mathbf{y}_{\rm p}}\mathbf{A}_{\rm p}^H,
\end{equation}
where the subscript `${\rm p}$' {represents either the training, uplink or downlink phase,} ${\rm p}\in\{{\rm t},{\rm u},{\rm d}\}$. According to the arcsine law, we can rewrite  $\mathbf{C}_{\bm{\eta}_{\rm p}}$ as
\begin{equation}\label{C_eta_eta}
\mathbf{C}_{\bm{\eta}_{\rm p}}  = \frac{2}{\pi}(\arcsin(\mathbf{X}_{\rm p})+j\arcsin(\mathbf{Y}_{\rm p})) - \frac{2}{\pi}(\mathbf{X}_{\rm p} + j\mathbf{Y}_{\rm p}),
\end{equation}
where
\begin{align}
\label{matrix_X}\mathbf{X}_{\rm p} &= \diag(\mathbf{C}_{\mathbf{y}_{\rm p}})^{-\frac{1}{2}} \Re\left({\mathbf{C}_{\mathbf{y}_{\rm p}}}\right)\diag(\mathbf{C}_{\mathbf{y}_{\rm p}})^{-\frac{1}{2}}, \\
\label{matrix_Y}\mathbf{Y}_{\rm p} &= \diag(\mathbf{C}_{\mathbf{y}_{\rm p}})^{-\frac{1}{2}} \Im\left({\mathbf{C}_{\mathbf{y}_{\rm p}}}\right)\diag(\mathbf{C}_{\mathbf{y}_{\rm p}})^{-\frac{1}{2}}.
\end{align}

{Due to the nonlinear arcsine function,} it is complicated to prove the SINR duality directly. However,
{we note the approximation that}
\begin{equation}\label{arcsin_approximation}
\frac{2}{\pi}\arcsin(a) \cong \left\{ \begin{array}{*{20}{c}}
{1,}&{a = 1}\\
{2a/\pi,}&{a<1.}
\end{array}
\right.
\end{equation}
Thus if the {non-diagonal} elements of $\mathbf{X}_{\rm p}$ and $\mathbf{Y}_{\rm p}$ are much smaller than 1, we can approximate \eqref{C_eta_eta} as
\begin{equation}\label{general_C_eta_app}
\mathbf{C}_{\bm{\eta}_{\rm p}} \cong (1-2/\pi)\mathbf{I},
\end{equation}
{which implies that the quantizer noise is approximately uncorrelated.} By using this approximation, we {arrive at Theorem 1 below}.

{\it Theorem 1}: {Assume a one-bit MIMO system with the uplink linear receiver $\mathbf{W}=[\mathbf{w}_1,...,\mathbf{w}_K]$, user power allocation vector $\mathbf{p} = [p_1,...,p_K]^T$ with $\|\mathbf{p}\|_1 = P_{\rm total}^{\rm (UL)}$ and uncorrelated quantizer noise. If the $k$th user achieves ${\rm SINR}_k^{\rm (UL)} = \gamma_k$ in the uplink, then the same ${\rm SINR}_k^{\rm (DL)} = \gamma_k$ can also be achieved using equal total power in the downlink if the downlink precoders and BS antenna power allocation is chosen as follows:
\begin{equation}
\mathbf{t}_i = \sqrt{\frac{q_i}{\|\mathbf{A}_{\rm u}\mathbf{w}_i\|_2^2}}\mathbf{A}_{\rm u}\mathbf{w}_i \triangleq \sqrt{q_i}\hat{\mathbf{t}}_i
\end{equation}
\begin{equation}\label{theorem1_power_allocation}
\mathbf{Q} = \diag\left(\sum_{i=1}^K \mathbf{t}_i\mathbf{t}_i^H\right)^{\frac{1}{2}},
\end{equation}
where $\mathbf{A}_{\rm u}$ is defined in \eqref{A_u} and  $q_i$ is the $i$th element of vector $\mathbf{q}$ defined by
\begin{equation}
\mathbf{q} = \frac{\pi}{2}(\mathbf{I} - \mathbf{D}\mathbf{\Psi})^{-1}\mathbf{D}\mathbf{1}_K,
\end{equation}
where $\|\mathbf{q}\|_1 = P_{\rm total}^{\rm (DL)} = P_{\rm total}^{\rm (UL)}$, $\mathbf{D}$ is a $K\times K$ diagonal matrix whose $k$th diagonal element is
\begin{equation}\label{DD}
  \left[\mathbf{D}\right]_{k,k} = \gamma_k\left|\mathbf{g}_k^T\hat{\mathbf{t}}_k\right|^{-2},
\end{equation}
and $\mathbf{\Psi}$ is a $K\times K$ matrix, whose $(k,i)$th element is
\begin{equation}\label{PsiPsi}
\left[\bm{\Psi}\right]_{k,i} = \left\{ \begin{aligned}
&{|\mathbf{g}_k^T\hat{\mathbf{t}}_i|^2 +  \left(\frac{\pi}{2}-1\right)\hat{\mathbf{t}}_i^H\diag(\mathbf{g}_k^*\mathbf{g}_k^T)\hat{\mathbf{t}}_i}&{i\neq k }\\
&{\left(\frac{\pi}{2}-1\right)\hat{\mathbf{t}}_i^H\diag(\mathbf{g}_k^*\mathbf{g}_k^T)\hat{\mathbf{t}}_i.}&{i = k}
\end{aligned}
\right.
\end{equation}}
\begin{IEEEproof}
See Appendix \ref{proof_duality}.
\end{IEEEproof}

{{\it Corollary 1}: Assume a one-bit MIMO system with the downlink linear precoder $\mathbf{T}=[\mathbf{t}_1,...,\mathbf{t}_K]$, BS antenna power allocation $\mathbf{Q} = \diag\left(\sum_{i=1}^K \mathbf{t}_i\mathbf{t}_i^H\right)^{\frac{1}{2}}$ with $\tr(\mathbf{Q}\mathbf{Q}^H) = P_{\rm total}^{\rm (DL)}$ and uncorrelated quantizer noise. If the $k$th user achieves the ${\rm SINR}_k^{\rm (DL)} = \gamma_k$ in the downlink, then the same ${\rm SINR}_k^{\rm (UL)} = \gamma_k$ can also be achieved using equal total power in the uplink if  the uplink receivers and user power allocation vector is chosen as
\begin{equation}
\mathbf{w}_i = \mathbf{A}_{\rm u}^{-1}{\mathbf{t}_i}/{\|\mathbf{t}_i\|_2}
\end{equation}
\begin{equation}
\mathbf{p}= \frac{\pi}{2}(\mathbf{I} - \mathbf{D}\mathbf{\Psi}^T)^{-1}\mathbf{D}\mathbf{1}_K,
\end{equation}
where $\|\mathbf{p}\|_1 = P_{\rm total}^{\rm (UL)} = P_{\rm total}^{\rm (DL)}$. The matrices $\mathbf{A}_{\rm u}$, $\mathbf{D}$ and $\mathbf{\Psi}$ are defined in \eqref{A_u}, \eqref{DD} and \eqref{PsiPsi}, respectively.}

Note that the uplink-downlink SINR duality holds for arbitrary user power allocation and linear receivers, and the key point is that the quantizer noise has to be uncorrelated. Since it is expected that the number of users is large and the SNR per-antenna is low due to the availability of a large array gain in massive MIMO systems, the assumption of uncorrelated quantizer noise using the approximation in \eqref{general_C_eta_app} is quite reasonable. Therefore, we conclude that uplink-downlink SINR duality holds in massive MIMO systems operating at low SNR.

\section{Achievable Rate Approximation in One-Bit Massive MIMO Systems}
In one-bit MIMO systems, the quantizer noise for the both uplink and downlink {is} not Gaussian due to the one-bit quantization. However the entropy of the uncorrelated effective noise is upper-bounded by the entropy of Gaussian noise, {and this minimizes} the mutual information of the input and output \cite{hassibi2003how}. Therefore, a lower bound for the achievable rate can be found by modeling the quantizer noise $\bm{\eta}_{\rm u}$ and $\bm{\eta}_{\rm d}$ as Gaussian with the same covariance matrix {given} in \eqref{quantizer_Covariance_Matrix}. Thus the ergodic achievable rate for both the uplink and downlink in one-bit MIMO systems is lower bounded by
\begin{equation}\label{ergodic_achievable_rate}
\tilde{R}_k^{\rm (Link)} = \E\left\{\log_2\left(1+{\rm SINR}_k^{\rm (Link)}\right)\right\},
\end{equation}
where ${\rm Link}\in\{\rm UL, DL\}$, and ${\rm SINR}_k^{\rm (UL)}$ and ${\rm SINR}_k^{\rm (DL)}$ are {given} in \eqref{SINR_UL} and \eqref{SINR_DL}, respectively \cite{ngo2013energy}.
In this section we {derive} closed-form expressions for the achievable rate in the low SNR region.

The covariance matrix of the estimated effective channel $\hat{\mathbf{g}}_k^{\rm eff}$ in {\eqref{channel_estimate} is given by}
\begin{equation}\label{cov_estimate_channel_exact}
\mathbf{C}_{\hat{\mathbf{g}}_k^{\rm eff}} = \rho_{\rm u}^2\tilde{\bm \phi}_k^H \mathbf{C}_{\mathbf{r}_{\rm t}} \tilde{\bm \phi}_k.
\end{equation}
We see that the expression above involves the covariance matrix $\mathbf{C}_{\mathbf{r}_{\rm t}}$, which is complicated due to the arcsine law of \eqref{C_rr}. However, using the approximation of \eqref{general_C_eta_app} {for low SNR} and using the linear form of $\mathbf{r}_{\rm t}$ {given} in \eqref{r_t_Bussgang}, the covariance matrix of $\mathbf{C}_{\bm{\eta}_{\rm t}}$ can be approximated as
\begin{align}
\mathbf{C}_{\mathbf{r}_{\rm t}} &= \sum_{k=1}^K \rho_{\rm u} \tilde{\bm{\phi}}_k\tilde{\bm{\phi}}_k^H + \mathbf{A}_{\rm t}\mathbf{A}_{\rm t}^H + \mathbf{C}_{\bm{\eta}_{\rm t}\bm{\eta}_{\rm t}} \nonumber \\
&\cong \sum_{k=1}^K \rho_{\rm u} \tilde{\bm{\phi}}_k\tilde{\bm{\phi}}_k^H + (\alpha^2+1-\frac{2}{\pi})\mathbf{I}.
\end{align}


Note that although the elements of the channel estimate \eqref{channel_estimate} are not Gaussian distributed due to the one-bit quantization, we can approximate them as Gaussian according to Cram{\' e}r's central limit theorem \cite{Cramer2004random} since each element of $\hat{\mathbf{g}}_k^{\rm eff}$ can be expressed as the summation of a large number of random variables. Therefore, in the sequel each element of the channel estimate $\hat{\mathbf{g}}_k^{\rm eff}$ {is} approximated as Gaussian with covariance matrix
\begin{align}
\mathbf{C}_{\hat{\mathbf{g}}_k^{\rm eff}} &\cong \rho_{\rm u}^2 \tilde{\bm{\phi}}_k^H\left(\sum_{k=1}^K \rho_{\rm u} \tilde{\bm{\phi}}_k\tilde{\bm{\phi}}_k^H + \left(\alpha^2+1-\frac{2}{\pi}\right)\mathbf{I}\right)^{-1}\tilde{\bm{\phi}}_k \nonumber \\
& = \frac{\alpha^2 \tau\rho_{\rm u}^2}{\alpha^2 \tau\rho_{\rm u} + \alpha^2+1-2/\pi}\mathbf{I} = \sigma^2\mathbf{I},
\end{align}
which implies that, {at low SNR}, each element of the channel estimate $\hat{\mathbf{g}}_k^{\rm eff}$ {is} approximated as independent and identically Gaussian distributed (i.i.d) with zero mean and variance $\sigma^2$.
\subsection{Uplink Transmission}
In this section we focus on deriving a closed-form expression {for the uplink achievable rate.}
Since there is no efficient way to directly calculate the achievable rate in \eqref{ergodic_achievable_rate}, we borrow Lemma 1 from \cite{zhang2014power} to derive an approximate closed-form expression. For completeness, we provide it below.

{\it Lemma 1}: If $X = \sum_{i=1}^{M}X_i$ and $Y = \sum_{i=j}^{M}Y_j$ are both sums of nonnegative random variables $X_i$ and $Y_j$, then the following approximation {holds}
\begin{equation}\label{E_Approximation}
\E\left\{\log_2\left(1+\frac{X}{Y}\right)\right\} \cong \log_2\left(1+\frac{\E\{X\}}{\E\{Y\}}\right).
\end{equation}
\begin{IEEEproof}
See \cite{zhang2014power}.
\end{IEEEproof}
Note that the approximation in Lemma 1 does not require the random variables $X$ and $Y$ to be independent and {becomes more accurate as $M$ increases.

According to Lemma 1 and substituting $\mathbf{A}_{\rm u} = \alpha\mathbf{I}$, we can approximate the uplink ergodic achievable rate $\tilde{R}_k^{\rm (UL)}$ by
\begin{equation}\label{ergodic_achievable_rate_App}
R_k^{\rm (UL)} = \log_2\left(1+\frac{\alpha^2\E\left\{\left|{\mathbf{{w}}}_k^T{{\bf{g}}_k^{\rm eff}}\right|^2\right\}}{\sum_{i \ne k}^K \alpha^2 \E\left\{\left|{{\mathbf{{w}}}_k^T {{\bf{g}}_i^{\rm eff}}}\right|^2\right\} + {\rm AQN}_k }\right),
\end{equation}
where we define
\begin{align}
\textrm{AQN}_k &= \alpha^2\E\left\{\left\|\mathbf{{w}}_k^T\right\|_2^2\right\} + \E\left\{\mathbf{{w}}_k^T \mathbf{C}_{\bm{\eta}_{\rm u}}\mathbf{w}_k^*\right\},
\end{align}
and the expectation operation is taken with respect to the channel realizations. In this paper, we will consider the performance of the {standard} MRC and ZF receivers, defined by
\begin{align}
&\mathbf{W}_{\rm MRC}^T= \hat{\mathbf{G}}_{\rm eff}^H \\
&\mathbf{W}_{\rm ZF}^T= \left(\hat{\mathbf{G}}_{\rm eff}^H \hat{\mathbf{G}}_{\rm eff}^H\right)^{-1}\hat{\mathbf{G}}_{\rm eff}^H
\end{align}
respectively, where $\hat{\mathbf{G}}_{\rm eff} = \left[\hat{\mathbf{g}}_1^{\rm eff},...,\hat{\mathbf{g}}_K^{\rm eff}\right]$.

{\it Theorem 2}: For an MRC receiver based on the LMMSE channel estimate and {operating at low SNR}, the uplink achievable rate for each active terminal in a one-bit massive MIMO system can be approximated by
\begin{equation}\label{MRC_closed_form}
R_{\rm MRC}^{\rm (UL)} = \log_2\left(1+\frac{\alpha^2(\sigma^2 M+\rho_{\rm u})}{\rho_{\rm u}\alpha^2(K-1)+\alpha^2+1-2/\pi}\right).
\end{equation}

\begin{IEEEproof}
See Appendix \ref{proof_MRC_achievable_rate}.
\end{IEEEproof}

{\it Theorem 3}: For a ZF receiver based on the LMMSE channel estimate and {operating at low SNR}, the uplink achievable rate {for} each active terminal in a one-bit massive MIMO system can be approximated by
\begin{equation}\label{ZF_closed_form}
R_{\rm ZF}^{\rm (UL)} = \log_2\left(1+\frac{\alpha^2\sigma^2(M-K -1 )+ \alpha^2 \rho_{\rm u}}{\alpha^2(K-1)(\rho_{\rm u}-\sigma^2)+\alpha^2+1-2/\pi}\right).
\end{equation}
\begin{IEEEproof}
See  Appendix \ref{proof_ZF_achievable_rate}.
\end{IEEEproof}

Note that equal achievable rates can be guaranteed to all active terminals due to the power control strategies; thus we omit the subscript $k$ in the achievable rate {expressions} of \eqref{MRC_closed_form} and \eqref{ZF_closed_form}.

\subsection{Downlink Transmission}
For conventional MIMO systems\footnote{{The term ``conventional MIMO system'' refers here to} MIMO systems equipped with infinite-resolution ADC/DACs.}, in order to satisfy the {conditions for} uplink-downlink duality, the precoding vectors are assumed to {satisfy} $\mathbf{t}_k = \sqrt{{q_k/\|\mathbf{w}_k\|^2}}\mathbf{w}_k$, where the scaling $q_k$ is chosen to satisfy the constraint that the total transmission power of the uplink and downlink are equal. However, according to Theorem 1, in order to ensure the SINR duality holds,
we consider {the modified} matched-filter (MF) and ZF precoding schemes {for the $k$th user
\begin{align}
\label{Preocding_matrix_MF}&\mathbf{t}_{k,{\rm MF}} = \sqrt{\frac{q_k}{\|\mathbf{A}_{\rm u}\mathbf{w}_{k,{\rm MRC}}\|^2}}\mathbf{A}_{\rm u}\mathbf{w}_{k,{\rm MRC}},  \\
\label{Preocding_matrix_ZF}&\mathbf{t}_{k,{\rm ZF}} = \sqrt{\frac{q_k}{\|\mathbf{A}_{\rm u}\mathbf{w}_{k,{\rm ZF}}\|^2}}\mathbf{A}_{\rm u}\mathbf{w}_{k,{\rm ZF}},
\end{align}
where $\mathbf{w}_{k,{\rm MRC}}$ and $\mathbf{w}_{k,{\rm ZF}}$ are the $k$th column of $\mathbf{W}_{{\rm MRC}}$ and $\mathbf{W}_{{\rm ZF}}$, respectively.}


{\it Theorem 4}: Employing the precoding strategies shown in \eqref{Preocding_matrix_MF} and \eqref{Preocding_matrix_ZF} and allocating transmit power as in \eqref{theorem1_power_allocation},
the downlink achievable rate of each active terminal {in one-bit MIMO systems} can be approximated by \eqref{MRC_closed_form} and \eqref{ZF_closed_form}, respectively.
\begin{IEEEproof}
This follows directly  from the uplink-downlink SINR duality in Theorem 1.
\end{IEEEproof}


\section{Optimal System Design of EE-SE Tradeoff \\for One-Bit Massive MIMO}
{Most prior work focuses on optimizing either the spectral or the energy efficiency {individually}.  Since the two are competing objectives, improving one inevitably leads to a degradation of the other. Since both are of importance for current wireless systems, here we investigate the joint optimization of both the} spectral and energy efficiency for one-bit massive MIMO systems by properly allocating {resources such as} the number of active terminals $K$, the training length $\tau$ and the system operating power $\rho_{\rm u}$ for a fixed number of BS antennas $M$ and {coherence} interval $T$.

The SE is defined as the average achievable rate over the {coherence} interval $T$:
\begin{align}\label{sum_SE_1}
\mathcal{F}_{{\rm SE}} =\frac{(T-\tau)}{T} K \left( \gamma R_{\rm A}^{\rm (UL)} + (1-\gamma) R_{\rm A}^{\rm (DL)}\right),
\end{align}
where ${\rm A}\in\{\rm MRC, ZF\}$. The EE is defined as {the} average achievable rate achieved per-unit transmit power:
\begin{equation}\label{Energy_Efficiency}
\mathcal{F}_{\rm EE} = \frac{(T-\tau)K}{T} \frac{ \gamma R_{\rm A}^{\rm (UL)} + (1-\gamma) R_{\rm A}^{\rm (DL)}}{\gamma \E\{P_{\rm total}^{\rm (UL)}\} + (1-\gamma) \E\{P_{\rm total}^{\rm (DL)}\}}.
\end{equation}

Note that the total transmit power $P_{\rm total}^{\rm (UL)}$ {changes} since the active terminals are randomly selected and the large scale fading coefficients are different {for each user}. Therefore, instead of considering the instantaneous total transmit power, here we consider the average total transmit power. For different channel realizations, the average total transmit power can be expressed as
\begin{align}\label{average_power}
\E\left\{P_{\rm total}^{\rm (UL)}\right\} &=\sum_{k=1}^K p_k =  \sum_{k=1}^K\E\{\beta_k^{-1}\} \rho_{\rm u}  \nonumber \\
& = K\rho_{\rm u} \frac{r_{\rm max}^{\kappa+2} - r_{\rm min}^{\kappa + 2}}{\bar{d}(1+\kappa/2)(r_{\rm max}^2-r_{\rm min}^2)r^{\kappa}_{\rm min}}.
\end{align}

{Since, according} to the uplink-downlink duality {derived} in Theorem~1, any target rate achieved in the uplink can be achieved in the downlink while {maintaining} $P_{\rm total}^{\rm(UL)}=P_{\rm total}^{\rm(DL)}$, we drop the superscript  `(UL)' and `(DL)' and {rewrite the SE and EE as}
\begin{align}
\label{sum_SE_2}\mathcal{F}_{\rm SE} &=\frac{(T-\tau)}{T} K  R_{\rm A}, \\
\label{Energy_Efficiency2}\mathcal{F}_{{\rm EE}} &= \frac{(T-\tau)}{T} \frac{ R_{\rm A} \bar{d}(1+\kappa/2)(r_{\rm max}^2-r_{\rm min}^2)r^{\kappa}_{\rm min}}{ \rho_{\rm u} (r_{\rm max}^{\kappa+2} - r_{\rm min}^{\kappa + 2})}.
\end{align}

Next we investigate the number of users and operating power that jointly optimize the spectral and energy efficiency  in one-bit massive MIMO systems. 
{Since the SE and EE are competing objectives, we will use a multiobjective optimization approach to balance the two criteria and obtain a Pareto optimal solution.  In particular, we will employ the weighted product method, expressed as}
\begin{equation}\label{product_weight_method}
(K^{\star}, \tau^{\star}, \rho_{\rm u}^{\star}) = \arg\mathop{\max}\limits_{K,\tau,\rho} ~~\left(\mathcal{F}_{\rm SE}\right)^{w_{\rm SE}}\cdot\left(  \mathcal{F}_{\rm EE}\right)^{w_{\rm EE}},
\end{equation}
where the weights ${w_{\rm SE}}$ and ${w_{\rm EE}}$ indicate the relative significance of the objective functions.
The specific Pareto optimal operating point  provided as the solution depends on which weights are used. For example, if $w_{\rm SE} \neq 0$ and $w_{\rm EE} = 0$, then the optimization problem reduces to {maximizing only} the spectral efficiency
{at the expense of} the energy efficiency. {The opposite is true} if $w_{\rm SE} = 0$ and $w_{\rm EE} \neq 0$. 
{The specific choice of the weights depends on the system design requirements. In the simulations that follow, we will show the Pareto boundary achieved by a wide range of weights, as well as the specific solution obtained with $w_{\rm SE} = w_{\rm EE} =1$.}

Note that, by using the uplink-downlink duality, it is guaranteed that the optimal solutions for the number of active terminals $K^{\star}$, training length $\tau^{\star}$ and the operating power $\rho_{\rm u}^{\star}$ is {the same for both the} uplink and downlink.

\section{Numerical Results}
For our simulations, we consider a cell with a radius of 500 meters and $K_{\max}$ terminals distributed randomly and uniformly over the cell, with the exclusion of a central disk of radius 100 meters. According to \cite{choi2008capacity}, we choose $\bar{d} = 10^{0.8}$ and $\kappa = 3.8$ for a typical urban cellular environment.

\begin{figure}
  \centering
  \includegraphics[width=9cm]{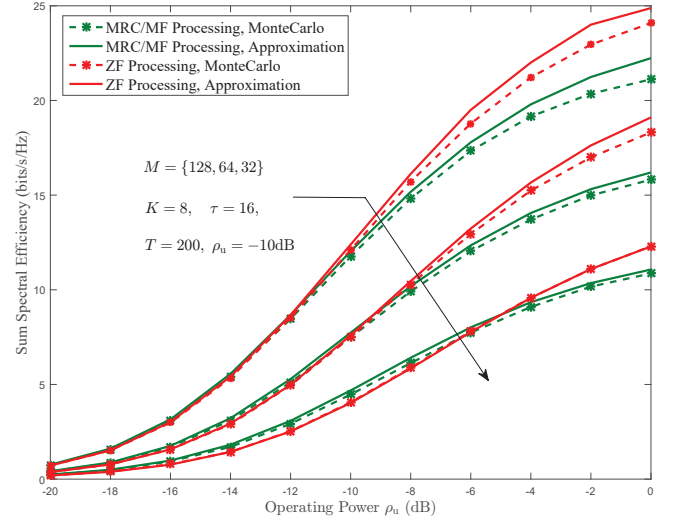}\\
  \vspace{-0.3cm}
  \caption{Sum spectral efficiency versus operating power $\rho_{\rm u}$ with $M = \{32, 64, 128\}$, $K = 8$, $T = 200$ and $\tau = 16$ for MRC/MF and ZF processing.}\label{MRC_ZF_MonteCarlo_App}
  \vspace{-0.5cm}
\end{figure}

We first evaluate the validity of the approximations for the achievable rate for the MRC and ZF receivers derived in Theorems 2 and 3 compared with the ergodic rate given in \eqref{ergodic_achievable_rate}. Fig.~\ref{MRC_ZF_MonteCarlo_App} shows the sum spectral efficiency versus operating power $\rho_{\rm u}$ for different numbers of transmit antennas $M = \{32, 64, 128\}$. We choose the length of the coherence interval to be $T = 200$, the number of active terminals $K = 8$, and training length $\tau = 16$, such that the relative pilot length $\tau_0 = 2$. The straight lines represent the sum spectral efficiencies obtained numerically from \eqref{MRC_closed_form} and \eqref{ZF_closed_form}, while the dashed lines with markers represent the ergodic sum spectral efficiencies obtained from \eqref{ergodic_achievable_rate}. Apparently, for both MRC and ZF processing, the gap between the bounds
and the empirical ergodic rates is small, especially {for low operating power}. This implies that the approximation {for}
the achievable rate given in \eqref{MRC_closed_form} and \eqref{ZF_closed_form} is a good predictor of the performance of one-bit massive MIMO systems. Thus, in the following plots we will show only the approximation when evaluating performance.

In the downlink, as analyzed in previous sections, since each transmit antenna should {be equipped} with a power amplifier in order to implement the power allocation in \eqref{theorem1_power_allocation}, it is {desirable} that the power amplifier {gains be confined to a limited range of values.}Fig.~\ref{DynamicPower} shows the cumulative density function of the downlink transmit power {of the individual BS antennas for} $M=\{32,64,128\}$ and total transmit power $P_{\rm total}^{\rm (DL)} = P_{\rm total}^{\rm (UL)}=10$dB for the modified MF and ZF precoding in \eqref{Preocding_matrix_MF} and \eqref{Preocding_matrix_ZF}, respectively. We assume the statistically-aware power control strategy is used in the uplink, while the power allocation as in \eqref{theorem1_power_allocation} is implemented in the downlink to ensure the SINR duality. We can see that the ranges of the power amplifier {gains} are indeed small. We also observe that the ranges for both the modified MF and ZF precoding decrease as the number of transmit antennas increases. Interestingly, we note that {for larger antenna arrays}, both modified MF and ZF precoding have almost the same ranges of power amplifier gains.

\begin{figure}
  \centering
  \includegraphics[width=9cm]{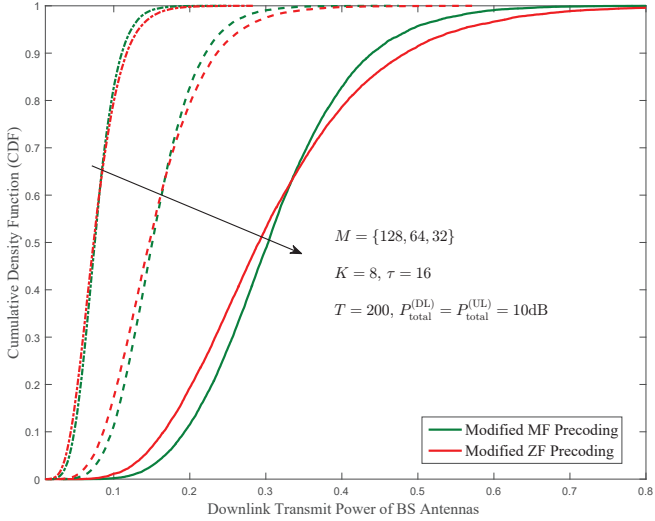}\\
  \vspace{-0.3cm}
  \caption{Cumulative density function of the downlink transmit power of BS antennas over 500 channel realizations for different numbers of transmit antennas $M$ and $P_{\rm total}^{\rm (DL)} = P_{\rm total}^{\rm (UL)} = 10$dB.}\label{DynamicPower}
  \vspace{-0.3cm}
\end{figure}

Next we study the SE vs. EE tradeoff. In conventional massive MIMO systems, a common rule of thumb is that the number of BS antennas $M$ should be roughly 10 times more than the number of active terminals $K$ and the optimal pilot training length should be equal to the number of active terminals, i.e., $\tau = K$. An interesting question is whether the aforementioned rule is still true for one-bit massive MIMO systems. To answer this question, {Fig.~\ref{MRC_ZF_Pareto_Boundary} plots the SE versus EE for} different numbers of BS antennas $M$ and a coherence interval $T = 400$ for MRC/MF and ZF processing. The solid lines and the dash-dot line represent the Pareto boundary of {the one-bit system design} and the unquantized case \cite{ngo2013energy}, respectively, when the number of active terminals $K$, the training length $\tau$ and the operating power $\rho_{\rm u}$ are jointly optimized. The dashed line represents the Pareto boundary of the benchmark case {for one-bit systems}, which is obtained by setting $M=200$, $K = 0.1M$, $\tau = K$ and only optimizing the operating power~$\rho_{\rm u}$.

\begin{figure}[t]

\subfloat[]{%
  \includegraphics[clip,width=\columnwidth]{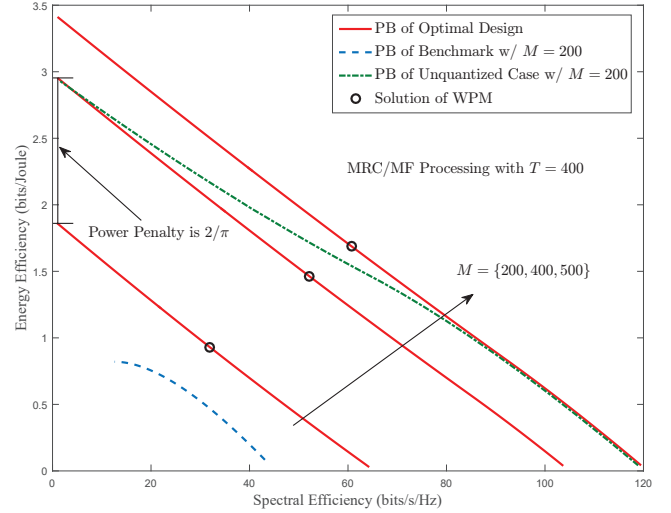}%
}

\subfloat[]{%
  \includegraphics[clip,width=\columnwidth]{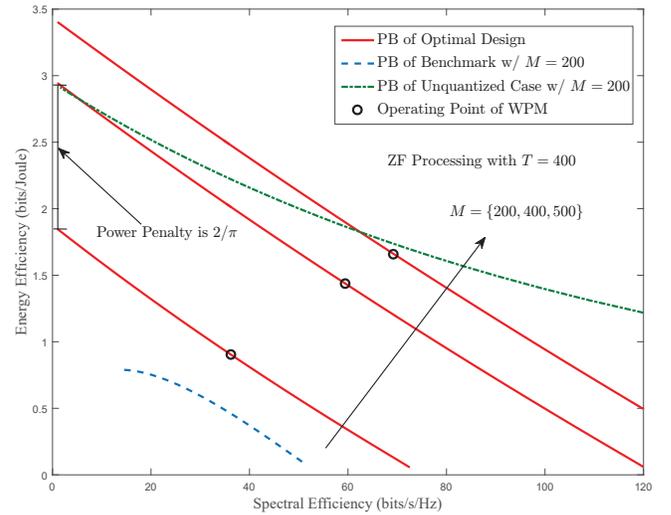}%
}

\caption{Pareto boundary of energy efficiency versus spectral efficiency for optimal design, benchmark and unquantized case with $M=\{200,400,500\}$ and $T = 400$ for (a) MRC/MF processing and (b) ZF processing.}\label{MRC_ZF_Pareto_Boundary}
\vspace{-0.3cm}
\end{figure}

%

{The tradeoff between SE and EE is clearly visible and in this example is nearly linear for the one-bit case; e.g., increasing the SE by 10\% leads to approximately a 10\% reduction in EE. We also note that the non-optimized benchmark configuration leads to a considerably lower operating point in terms of both SE and EE; optimization of the system parameters provides a significant benefit.  {In both figures, we see the power penalty of $2/\pi$ that separates the performance of conventional unquantized massive MIMO from that for one-bit systems at low SNR (low SE)}. The curves for the conventional unquantized system plotted for $M=200$ show different behavior for the MRC/MF and ZF methods. We see that for MRC/MF processing, the optimized one-bit system can achieve the same performance as the unquantized system when the number of antennas deployed ranges from twice as many ($M=400$) at low SE values (low operating SNR) to 2.5 times as many ($M=500$) for large SEs (high operating SNR). For ZF processing, twice as many antennas are also needed for low SNR/SE, but the amount of extra antennas grows beyond a factor of 2.5 for high SNR/SE. This is because at high SNR, unlike MRC/MF, the quality of the estimated CSI in the conventional system improves and ZF can eliminate more and more multiuser interference. The operating point achieved by \eqref{product_weight_method} with $w_{\rm SE}=w_{\rm EE}=1$ shown by the small circle on each Pareto boundary yields a reasonable compromise between the SE and EE.}

\begin{figure}
  \centering
  \includegraphics[width=9cm]{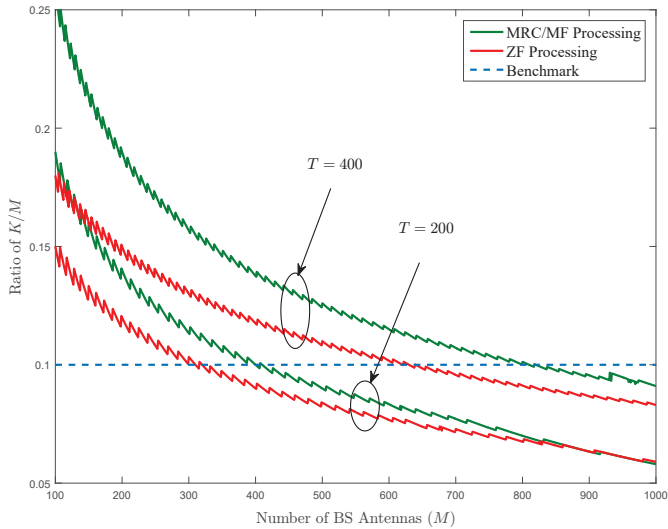}\\
  \vspace{-0.3cm}
  \caption{{Optimal ratio} of $K/M$ versus the number of BS antennas $M$ {for} different {coherence} intervals $T$ for MRC/MF and ZF processing.}\label{WPM_K_M_vs_M}
  \vspace{-0.5cm}
\end{figure}

\begin{figure}[!t]
  \centering
  \includegraphics[width=9cm]{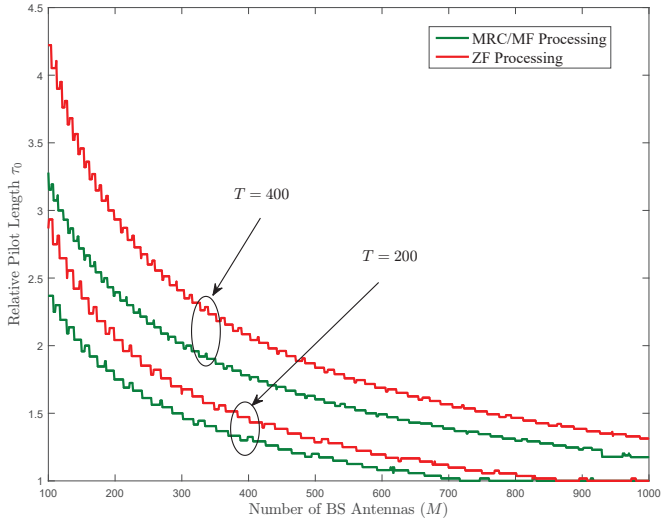}\\
  \vspace{-0.3cm}
  \caption{Optimal relative pilot length $\tau_0$ versus the number of BS antennas $M$ for different {coherence} intervals $T$ for MRC/MF and ZF processing.}\label{WPM_tau0_vs_M}
  \vspace{-0.3cm}
\end{figure}

Fig.~\ref{WPM_K_M_vs_M}-\ref{WPM_tau0_vs_M} show the {optimal} ratio of $K^{\star}/M$ and the {optimal} relative training length $\tau_0$, {which are obtained by \eqref{product_weight_method} with $w_{\rm SE} = w_{\rm EE} = 1$}, versus the number of BS antennas $M$ for different coherence intervals $T$. We see from the plots  that the optimal number of active terminals and pilot training length {does not generally follow} the assumption of $K=0.1M$ and $\tau = K$. 
{The optimal loading factor $K/M$ can be as high as 0.25 for MRC/MF when $M=100$ and $T=400$, but it decreases as $M$ increases or $T$ decreases.  The optimal length of the training interval is considerably larger for one-bit systems compared with conventional MIMO, where $\tau_0=1$ is known to be optimal. For values of $M$ below 200, 2-4 times as much training should be used depending on $T$ and the type of processing used. For larger and larger $M$, $\tau_0$ converges towards unity.}

Since our analysis {is based on the assumption of} low operating power, we finally verify whether this assumption is reasonable or not. Fig.~\ref{WPM_rho_vs_M} illustrates the {optimal} operating power $\rho_{\rm u}$ versus the number of BS antennas $M$ for MRC/MF and ZF processing for different coherence intervals. {We see from the curves that for $M$ and $T$ larger than 100, the optimal operating SNR is less than -9 dB,} which implies that our low operating power assumption is {justified}. Combining {the observations from} Fig.~\ref{WPM_K_M_vs_M} with Fig.~\ref{WPM_rho_vs_M}, it is interesting to note that, {when one considers EE together with SE,} MRC/MF processing {competes favorably with} ZF processing since it uses less power while serving more terminals with lower computational complexity.

\begin{figure}
  \centering
  \includegraphics[width=9cm]{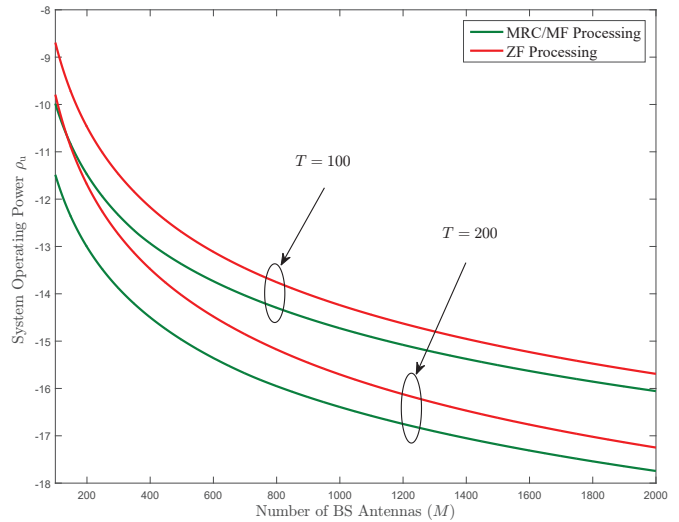}\\
  \vspace{-0.3cm}
  \caption{Optimal system operating power $\rho_{\rm u}$ versus the number of BS antennas $M$ {for} different {coherence} intervals $T$ for MRC/MF and ZF processing.}\label{WPM_rho_vs_M}
  \vspace{-0.5cm}
\end{figure}

\section{Conclusions}
This paper has investigated the optimal design {of} single-cell massive MIMO systems {employing one-bit} ADC/DACs at the BS. {Using the Bussgang decomposition, we derived approximate closed-form expressions at low SNR for the uplink achievable rate in a one-bit massive MIMO system assuming MRC or ZF receivers that operate with imperfect CSI. We derived the conditions under which uplink-downlink SINR duality holds for one-bit massive MIMO systems, and under these conditions we examined the spectral and energy efficiency tradeoff. We optimized a weighted product of the two competing objectives over the number of active users, the length of the training data and the operating power and compared the resulting performance with an unoptimized benchmark system design and the optimized unquantized massive MIMO system. The results show that depending on the operating SNR, the one-bit systems require 2-2.5 times more antennas to achieve the same spectral/energy efficiency as a conventional unquantized system. We also illustrated that, in general, one-bit systems require more training than conventional systems at the optimal operating point, up to 2.5 times as much depending on the length of the coherence interval and the number of antennas. We also observed that MRC/MF processing can be competitive when compared with ZF when energy efficiency is taken into account, since it can service more users with less operating power and computational complexity.}

\appendices
\section{}\label{proof_duality}
Suppose $\gamma_k$ is the target SINR achieved by the $k$th active user for both uplink and downlink transmission. For the downlink, we assume the precoding vector is
\begin{equation}
\mathbf{t}_k = {\sqrt{q_k} \frac{\mathbf{A}_{\rm u}\mathbf{w}_k}{\|\mathbf{A}_{\rm u}\mathbf{w}_k\|_2}} \triangleq \sqrt{q_k}\hat{\mathbf{t}}_k.
\end{equation}
where $q_k = \|\mathbf{t}_k\|^2_2$, {$\hat{\mathbf{t}}_k = {{\mathbf{t}}_k}/\sqrt{q_k}$ with}  $\|\hat{\mathbf{t}}_k\|^2_2 = 1$, and the power allocation for the downlink is
\begin{equation}\label{Power_Matrix_in_Appendix}
\mathbf{Q} = {\diag\left(\sum_{i=1}^K \mathbf{t}_i\mathbf{t}_i^H\right)^{\frac{1}{2}}}.
\end{equation}
With this notation, we have
\begin{equation}
P_{\rm total}^{\rm (DL)} = {\tr(\mathbf{Q}\mathbf{Q}^H) = \sum_{i=1}^K q_i,}
\end{equation}
which allows us to transfer the problem of power allocation for each antenna into the problem of power allocation for each precoding vector. Substituting {\eqref{Power_Matrix_in_Appendix}} into \eqref{SINR_DL} and assuming uncorrelated quantizer noise $\bm{\eta}_{\rm d}$ such that $\mathbf{C}_{\bm{\eta}_{\rm d}} = (1-\frac{2}{\pi})\mathbf{I}$, the downlink SINR can be expressed as
\begin{align}\label{downlink_duality}
&\gamma_k = \frac{q_k|\mathbf{g}_k^T\hat{\mathbf{t}}_k|^2}{{\sum_{i\neq k}^K q_i |\mathbf{g}_k^T\hat{\mathbf{t}}_i|^2 + (\frac{\pi}{2} - 1) \mathbf{g}_k^T\diag(\sum_{i=1}^K q_i\hat{\mathbf{t}}_i\hat{\mathbf{t}}_i^H)\mathbf{g}_k^* + {\frac{\pi}{2}}}} \nonumber \\
&=\frac{q_k|\mathbf{g}_k^T\hat{\mathbf{t}}_k|^2}{{\sum_{i\neq k}^K q_i |\mathbf{g}_k^T\hat{\mathbf{t}}_i|^2 + (\frac{\pi}{2} - 1) \sum_{i=1}^K q_i\hat{\mathbf{t}}_i^H\diag(\mathbf{g}_k^*\mathbf{g}_k^T)\hat{\mathbf{t}}_i + {\frac{\pi}{2}}}}.
\end{align}
The second line in \eqref{downlink_duality} holds thanks to the following identity for two matrices $\mathbf{A}$, $\mathbf{B}$:
\begin{equation}\label{trace_diag_identity}
\tr(\mathbf{A}\diag(\mathbf{B})) = \tr(\mathbf{B}\diag(\mathbf{A})).
\end{equation}
{Then by using the definitions in \eqref{DD} and \eqref{PsiPsi},
\eqref{downlink_duality} can be rewritten as}
\begin{equation}\label{downlink_SINR_condition}
\left[\mathbf{D}\right]_{k,k} = \frac{q_k}{\sum_{i=1}^K q_i\left[\bm{\Psi}\right]_{k,i}+{\frac{\pi}{2}}}.
\end{equation}
Therefore, according to \eqref{downlink_SINR_condition} we can express the downlink SINR conditions for all active users as
\begin{equation}
\mathbf{q} = \mathbf{D}\mathbf{\Psi}\mathbf{q} + \frac{\pi}{2}\mathbf{D}\mathbf{1}_K \Leftrightarrow \mathbf{q} = {\frac{\pi}{2}}(\mathbf{I} - \mathbf{D}\mathbf{\Psi})^{-1}\mathbf{D}\mathbf{1}_K,
\end{equation}
where $\mathbf{q} = [q_1,...,q_k]^T$ and $\mathbf{1}_K$ denotes the all-one vector.

For the uplink transmission, {since $\frac{\mathbf{A}_{\rm u}\mathbf{w}_k}{\|\mathbf{A}_{\rm u}\mathbf{w}_k\|_2} = \hat{\mathbf{t}}_k$,} the uplink SINR in \eqref{SINR_UL} can be rewritten as
\begin{align}\label{uplink_duality}
\gamma_k &= \frac{p_k |\hat{\mathbf{t}}_k^T{{\bf{g}}_k}|^2}{\sum_{i \ne k}^K p_i |{\hat{\mathbf{t}}_k^T{{\bf{g}}_i}}|^2  + 1 + (1-\frac{2}{\pi}) \hat{\mathbf{t}}_k^T \mathbf{A}_{\rm u}^{-2}\hat{\mathbf{t}}_k^*}.
\end{align}
After substituting \eqref{A_u} into \eqref{uplink_duality}, the denominator of \eqref{uplink_duality} can be expressed as
\begin{align}\label{uplink_duality_denominator}
&\sum_{i \ne k}^K p_i |{\hat{\mathbf{t}}_k^T{{\bf{g}}_i}}|^2  + 1 + \left(\frac{\pi}{2}-1\right) \hat{\mathbf{t}}_k^T \diag\left(\sum_{i=1}^K p_i \mathbf{g}_i\mathbf{g}_i^H + \mathbf{I}\right)\hat{\mathbf{t}}_k^* \nonumber\\
& = \sum_{i \ne k}^K p_i |{\hat{\mathbf{t}}_k^T{{\bf{g}}_i}}|^2  + \left(\frac{\pi}{2}-1\right)\sum_{i=1}^K p_i \hat{\mathbf{t}}_k^H\diag\left(\mathbf{g}_i^*\mathbf{g}_i^T  \right)\hat{\mathbf{t}}_k + \frac{\pi}{2}.
\end{align}
By using the notation of \eqref{DD} and \eqref{PsiPsi}, we can then express \eqref{uplink_duality} as
\begin{equation}\label{downlink_SINR_condition}
\left[\mathbf{D}\right]_{k,k} = \frac{p_k}{\sum_{i=1}^K p_i\left[\bm{\Psi}\right]_{i,k}+\frac{\pi}{2}}.
\end{equation}
After stacking it into vector form, we have
\begin{equation}
\mathbf{p} = \mathbf{D}\mathbf{\Psi}^T\mathbf{p} + \frac{\pi}{2}\mathbf{D}\mathbf{1}_K \Leftrightarrow
\mathbf{p} = \frac{\pi}{2}(\mathbf{I}-\mathbf{D}\mathbf{\Psi}^T)^{-1}\mathbf{D}\mathbf{1}_K,
\end{equation}
where $\mathbf{p} = [p_1,...,p_K]^T$. {Using the matrix inversion lemma and the fact that $\mathbf{D} = \mathbf{D}^T$, we have $(\mathbf{I}-\mathbf{D}\mathbf{\Psi}^T)^{-1}\mathbf{D} = ((\mathbf{I}-\mathbf{D}\mathbf{\Psi})^{-1}\mathbf{D})^T$, which implies that the total transmit power for the uplink and downlink are the same:}
\begin{equation}
{P_{\rm total}^{\rm (UL)} = \|\mathbf{p}\|_1 = \|\mathbf{q}\|_1 = P_{\rm total}^{\rm (DL)}}.
\end{equation}
This completes the proof.

\section{}\label{proof_MRC_achievable_rate}
From \eqref{ergodic_achievable_rate_App}, we first focus on deriving the terms $\E\left\{\left|{\mathbf{{w}}}_k^T{{\bf{g}}_k^{\rm eff}}\right|^2\right\}$ and $\E\left\{\left|{{\mathbf{{w}}}_k^T {{\bf{g}}_i^{\rm eff}}}\right|^2\right\}$. For the MRC receiver, we have
\begin{equation}
\mathbf{w}_k^T \mathbf{g}_k^{\rm eff} = (\hat{\mathbf{g}}_k^{\rm eff})^H \hat{\mathbf{g}}_k^{\rm eff} + (\hat{\mathbf{g}}_k^{\rm eff})^H \bm{\varepsilon}_k.
\end{equation}
Since the elements of the estimated effective channel $(\hat{\mathbf{g}}_k^{\rm eff})$ are i.i.d, according to \cite{tulino2004random} we can thus obtain
\begin{align}\label{MRC_E_second_order}
&\E\left\{\left|{\mathbf{{w}}}_k^T{{\bf{g}}_k^{\rm eff}}\right|^2\right\} = \E\left\{\left|(\hat{\mathbf{g}}_k^{\rm eff})^H \hat{\mathbf{g}}_k^{\rm eff}\right|^2\right\} + \E\left\{\left|(\hat{\mathbf{g}}_k^{\rm eff})^H \bm{\varepsilon}_k\right|^2\right\} \nonumber \\
& = \sigma^4(M^2+M) + \sigma^2(\rho_{\rm u}-\sigma^2)M = \sigma^2M (\sigma^2M + \rho_{\rm u}).
\end{align}
For $i\neq k$, we have
\begin{equation}
\E\left\{\left|{(\hat{\mathbf{g}}_k^{\rm eff})^H {{\bf{g}}_i^{\rm eff}}}\right|^2\right\} =\Var\left\{{(\hat{\mathbf{g}}_k^{\rm eff})^H {{\bf{g}}_i^{\rm eff}}}\right\}= \rho_{\rm u} \sigma^2 M.
\end{equation}

Next we focus on the term ${\rm AQN}_k$.
{Using \eqref{general_C_eta_app},} we rewrite ${\rm AQN}_k$ as
\begin{equation}
{\rm AQN}_k = (\alpha_{\rm u}^2 + 1-2/\pi) \E\left\{\left\|\mathbf{{w}}_k^T\right\|_2^2\right\}.
\end{equation}
With the MRC receiver $\mathbf{w}_k^T = (\hat{\mathbf{g}}_k^{\rm eff})^H$, we have
\begin{equation}\label{MRC_AQN}
{\rm AQN}_k = (\alpha_{\rm u}^2 + 1-2/\pi) \sigma^2 M.
\end{equation}
Substituting \eqref{MRC_E_second_order}-\eqref{MRC_AQN} into \eqref{ergodic_achievable_rate_App}, we can arrive at Theorem~2.

\section{}\label{proof_ZF_achievable_rate}

For the ZF receiver, we have
\begin{equation}
\mathbf{W}_{\rm ZF}^T \mathbf{G}_{\rm eff} = \mathbf{W}_{\rm ZF}^T (\hat{\mathbf{G}}_{\rm eff} + \mathbf{\mathcal{E}}_{\rm eff}) = \mathbf{I} + \mathbf{W}_{\rm ZF}^T \mathbf{\mathcal{E}}_{\rm eff},
\end{equation}
where $\mathbf{\mathcal{E}}_{\rm eff} = [\bm{\varepsilon}_1^{\rm eff},...,\bm{\varepsilon}_K^{\rm eff}]$. Therefore
\begin{equation}
\mathbf{w}_k^T \mathbf{g}_k^{\rm eff} = 1 + \mathbf{w}_k^T\bm{\varepsilon}_k^{\rm eff},
\end{equation}
and $\E\left\{\left|{\mathbf{{w}}}_k^T{{\bf{g}}_k^{\rm eff}}\right|^2\right\}$ can be expressed as
\begin{align}\label{ZF_E_second_order}
\E&\left\{\left|{\mathbf{{w}}}_k^T{{\bf{g}}_k^{\rm eff}}\right|^2\right\}= 1 + \E\{|\mathbf{w}_k^T\bm{\varepsilon}_k^{\rm eff}|^2\} \nonumber \\
& = 1 + (\rho_{\rm u}-\sigma^2)\E\left\{\left[\left(\hat{\mathbf{G}}_{\rm eff}^H\hat{\mathbf{G}}_{\rm eff}\right)^{-1}\right]_{k,k}\right\}.
\end{align}
Since the elements of $\hat{\mathbf{G}}^{\rm eff}$ are approximated as i.i.d Gaussian, $(\hat{\mathbf{G}}^{\rm eff})^H\hat{\mathbf{G}}^{\rm eff}$ is a $K\times K $ central Wishart matrix with $M$ degrees of freedom. Thus
\begin{equation}
\E\left\{\left|{\mathbf{{w}}}_k^T{{\bf{g}}_k^{\rm eff}}\right|^2\right\} = 1+ \frac{\rho_{\rm u}-\sigma^2}{\sigma^2(M-K)}.
\end{equation}

For $i\neq k$, we have
\begin{align}
\mathbf{w}_k^T \mathbf{g}_i^{\rm eff}  &= \E\{|\mathbf{w}_k^T\bm{\varepsilon}_i^{\rm eff}|^2\} \nonumber \\
&= (\rho_{\rm u}-\sigma^2)\E\left\{\left[\left(\hat{\mathbf{G}}_{\rm eff}^H\hat{\mathbf{G}}_{\rm eff}\right)^{-1}\right]_{k,k}\right\} \nonumber \\
&= \frac{\rho_{\rm u}-\sigma^2}{\sigma^2(M-K)}.
\end{align}

Similarly, we can obtain
\begin{equation}\label{ZF_AQN}
{\rm AQN}_k = \frac{\alpha_{\rm u}^2 + 1-2/\pi }{\zeta_k^2 (M-K)}.
\end{equation}

Substituting \eqref{ZF_E_second_order}-\eqref{ZF_AQN} into \eqref{ergodic_achievable_rate_App}, we arrive at Theorem~3.

\bibliographystyle{IEEEtran}
\bibliography{reference}

\begin{thebibliography}{10}
\providecommand{\url}[1]{#1}
\csname url@samestyle\endcsname
\providecommand{\newblock}{\relax}
\providecommand{\bibinfo}[2]{#2}
\providecommand{\BIBentrySTDinterwordspacing}{\spaceskip=0pt\relax}
\providecommand{\BIBentryALTinterwordstretchfactor}{4}
\providecommand{\BIBentryALTinterwordspacing}{\spaceskip=\fontdimen2\font plus
\BIBentryALTinterwordstretchfactor\fontdimen3\font minus
  \fontdimen4\font\relax}
\providecommand{\BIBforeignlanguage}[2]{{%
\expandafter\ifx\csname l@#1\endcsname\relax
\typeout{** WARNING: IEEEtran.bst: No hyphenation pattern has been}%
\typeout{** loaded for the language `#1'. Using the pattern for}%
\typeout{** the default language instead.}%
\else
\language=\csname l@#1\endcsname
\fi
#2}}
\providecommand{\BIBdecl}{\relax}
\BIBdecl

\bibitem{marzetta2010noncooperative}
T.~Marzetta, ``Noncooperative cellular wireless with unlimited numbers of base
  station antennas,'' \emph{IEEE Transactions on Wireless Communications},
  vol.~9, no.~11, pp. 3590--3600, November 2010.

\bibitem{lu2014overview}
L.~Lu, G.~Y. Li, A.~L. Swindlehurst, A.~Ashikhmin, and R.~Zhang, ``An overview
  of massive {{MIMO}}: Benefits and challenges,'' \emph{IEEE Journal of
  Selected Topics in Signal Processing}, vol.~8, no.~5, pp. 742--758, 2014.

\bibitem{larsson2013massive}
E.~Larsson, O.~Edfors, F.~Tufvesson, and T.~Marzetta, ``Massive {{MIMO}} for
  next generation wireless systems,'' \emph{IEEE Communications Magazine},
  vol.~52, no.~2, pp. 186--195, February 2014.

\bibitem{rusek2013scaling}
F.~Rusek, D.~Persson, B.~K. Lau, E.~Larsson, T.~Marzetta, O.~Edfors, and
  F.~Tufvesson, ``Scaling up {{MIMO}}: Opportunities and challenges with very
  large arrays,'' \emph{IEEE Signal Processing Magazine}, vol.~30, no.~1, pp.
  40--60, Jan 2013.

\bibitem{murmann2016adc}
\BIBentryALTinterwordspacing
B.~Murmann, ``{ADC} performance survey 1997-2016.'' [Online]. Available:
  \url{http://web.stanford.edu/~murmann/adcsurvey.html}
\BIBentrySTDinterwordspacing

\bibitem{chiara2014massive}
C.~Risi, D.~Persson, and E.~G. Larsson, ``Massive {{MIMO}} with 1-bit {ADC},''
  \emph{Mathematics}, 2014.

\bibitem{juncil2015near}
J.~Choi, J.~Mo, and R.~W. Heath, ``Near maximum-likelihood detector and channel
  estimator for uplink multiuser massive {MIMO} systems with one-bit {ADC}s,''
  \emph{IEEE Transactions on Communications}, vol.~64, no.~5, pp. 2005--2018,
  May 2016.

\bibitem{yongzhi2016channel}
\BIBentryALTinterwordspacing
Y.~Li, C.~Tao, L.~Liu, A.~Mezghani, G.~Seco-Granados, and A.~Swindlehurst,
  ``Channel estimation and performance analysis in one-bit massive {MIMO}
  systems.'' [Online]. Available: \url{http://arxiv.org/abs/1504.03516}
\BIBentrySTDinterwordspacing

\bibitem{mollen2016performance}
C.~Moll\'{e}n, J.~Choi, E.~G. Larsson, and R.~W. {Heath~Jr.}, ``Uplink
  performance of wideband massive {MIMO} with one-bit {ADC}s,'' \emph{IEEE
  Transactions on Wireless Communications}, vol.~16, no.~1, pp. 87--100, 2017.

\bibitem{mezghani2008analysis}
A.~Mezghani and J.~Nossek, ``Analysis of {Rayleigh}-fading channels with 1-bit
  quantized output,'' in \emph{IEEE International Symposium on Information
  Theory (ISIT)}, July 2008, pp. 260--264.

\bibitem{nossek2006capacity}
J.~A. Nossek and M.~T. Ivrla{\v{c}}, ``Capacity and coding for quantized
  {{MIMO}} systems,'' in \emph{Proceedings of the international conference on
  Wireless communications and mobile computing}.\hskip 1em plus 0.5em minus
  0.4em\relax ACM, 2006, pp. 1387--1392.

\bibitem{emil2016massive}
E.~Bj{\" o}rnson, E.~G. Larsson, and M.~Debbah, ``Massive {MIMO} for maximal
  spectral efficiency: How many users and pilots should be allocated?''
  \emph{IEEE Transactions on Wireless Communications}, vol.~15, no.~2, pp.
  1293--1308, Feb 2016.

\bibitem{ngo2014massive}
H.~Q. Ngo, M.~Matthaiou, and E.~G. Larsson, ``Massive {MIMO} with optimal power
  and training duration allocation,'' \emph{IEEE Wireless Communications
  Letters}, vol.~3, no.~6, pp. 605--608, Dec 2014.

\bibitem{han2011green}
C.~Han, T.~Harrold, S.~Armour, I.~Krikidis, S.~Videv, P.~M. Grant, H.~Haas,
  J.~S. Thompson, I.~Ku, C.~X. Wang, T.~A. Le, M.~R. Nakhai, J.~Zhang, and
  L.~Hanzo, ``Green radio: Radio techniques to enable energy-efficient wireless
  networks,'' \emph{IEEE Communications Magazine}, vol.~49, no.~6, pp. 46--54,
  June 2011.

\bibitem{chen2011fundamental}
Y.~Chen, S.~Zhang, S.~Xu, and G.~Y. Li, ``Fundamental trade-offs on green
  wireless networks,'' \emph{IEEE Communications Magazine}, vol.~49, no.~6, pp.
  30--37, June 2011.

\bibitem{ngo2013energy}
H.~Q. Ngo, E.~Larsson, and T.~Marzetta, ``Energy and spectral efficiency of
  very large multiuser {{MIMO}} systems,'' \emph{IEEE Transactions on
  Communications}, vol.~61, no.~4, pp. 1436--1449, April 2013.

\bibitem{emil2015optimal}
E.~Bj{\" o}rnson, L.~Sanguinetti, J.~Hoydis, and M.~Debbah, ``Optimal design of
  energy-efficient multi-user {MIMO} systems: Is massive {MIMO} the answer?''
  \emph{IEEE Transactions on Wireless Communications}, vol.~14, no.~6, pp.
  3059--3075, June 2015.

\bibitem{marler2004survey}
R.~T. Marler and J.~S. Arora, ``Survey of multi-objective optimization methods
  for engineering,'' \emph{Structural and multidisciplinary optimization},
  vol.~26, no.~6, pp. 369--395, 2004.

\bibitem{branke2008multiobjective}
J.~Branke, K.~Deb, K.~Miettinen, and R.~Slowi{\'n}ski, \emph{Multiobjective
  optimization: interactive and evolutionary approaches}.\hskip 1em plus 0.5em
  minus 0.4em\relax Springer, 2008, vol. 5252.

\bibitem{bjornson2014multiobjective}
E.~Bj{\" o}rnson, E.~A. Jorswieck, M.~Debbah, and B.~Ottersten,
  ``Multiobjective signal processing optimization: The way to balance
  conflicting metrics in {5G} systems,'' \emph{IEEE Signal Processing
  Magazine}, vol.~31, no.~6, pp. 14--23, 2014.

\bibitem{tang2014resource}
J.~Tang, D.~K.~C. So, E.~Alsusa, and K.~A. Hamdi, ``Resource efficiency: A new
  paradigm on energy efficiency and spectral efficiency tradeoff,'' \emph{IEEE
  Transactions on Wireless Communications}, vol.~13, no.~8, pp. 4656--4669, Aug
  2014.

\bibitem{he2013energy}
C.~He, B.~Sheng, P.~Zhu, X.~You, and G.~Y. Li, ``Energy- and
  spectral-efficiency tradeoff for distributed antenna systems with
  proportional fairness,'' \emph{IEEE Journal on Selected Areas in
  Communications}, vol.~31, no.~5, pp. 894--902, May 2013.

\bibitem{zhang2016spectral}
Z.~Zhang, Z.~Chen, M.~Shen, and B.~Xia, ``Spectral and energy efficiency of
  multipair two-way full-duplex relay systems with massive {MIMO},'' \emph{IEEE
  Journal on Selected Areas in Communications}, vol.~34, no.~4, pp. 848--863,
  April 2016.

\bibitem{hao2015energy}
Y.~Hao, Z.~Song, S.~Hou, and H.~Li, ``Energy- and spectral-efficiency tradeoff
  in massive {MIMO} systems with inter-user interference,'' in \emph{IEEE 26th
  Annual International Symposium on Personal, Indoor, and Mobile Radio
  Communications (PIMRC)}, Aug 2015, pp. 553--557.

\bibitem{bussgang1952yq}
J.~J. Bussgang, ``Crosscorrelation functions of amplitude-distorted {Gaussian}
  signals,'' MIT Research Lab. Electronics, Tech. Rep. 216, 1952.

\bibitem{biguesh2004downlink}
M.~Biguesh and A.~B. Gershman, ``Downlink channel estimation in cellular
  systems with antenna arrays at base stations using channel probing with
  feedback,'' \emph{EURASIP Journal on Applied Signal Processing}, vol. 2004,
  pp. 1330--1339, 2004.

\bibitem{bar2002doa}
O.~Bar-Shalom and A.~Weiss, ``{DOA} estimation using one-bit quantized
  measurements,'' \emph{IEEE Transactions on Aerospace and Electronic Systems},
  vol.~38, no.~3, pp. 868--884, 2002.

\bibitem{kay1993fundamentals}
S.~M. Kay, \emph{Fundamentals of statistical signal processing: Estimation
  Theory}.\hskip 1em plus 0.5em minus 0.4em\relax Upper Saddle River, NJ, USA:
  Prentice Hall, 1993.

\bibitem{jacovitti1994estimation}
G.~Jacovitti and A.~Neri, ``Estimation of the autocorrelation function of
  complex {Gaussian} stationary processes by amplitude clipped signals,''
  \emph{IEEE Transactions on Information Theory}, vol.~40, no.~1, pp. 239--245,
  Jan 1994.

\bibitem{vishwanath2003duality}
S.~Vishwanath, N.~Jindal, and A.~Goldsmith, ``Duality, achievable rates, and
  sum-rate capacity of {Gaussian} {MIMO} broadcast channels,'' \emph{IEEE
  Transactions on Information Theory}, vol.~49, no.~10, pp. 2658--2668, 2003.

\bibitem{hassibi2003how}
B.~Hassibi and B.~Hochwald, ``How much training is needed in multiple-antenna
  wireless links?'' \emph{IEEE Transactions on Information Theory}, vol.~49,
  no.~4, pp. 951--963, April 2003.

\bibitem{Cramer2004random}
H.~Cram{\'e}r, \emph{Random variables and probability distributions}.\hskip 1em
  plus 0.5em minus 0.4em\relax Cambridge University Press, 2004, vol.~36.

\bibitem{zhang2014power}
Q.~Zhang, S.~Jin, K.-K. Wong, H.~Zhu, and M.~Matthaiou, ``Power scaling of
  uplink massive {MIMO} systems with arbitrary-rank channel means,'' \emph{IEEE
  Journal of Selected Topics in Signal Processing}, vol.~8, no.~5, pp.
  966--981, Oct 2014.

\bibitem{choi2008capacity}
W.~Choi and J.~G. Andrews, ``The capacity gain from intercell scheduling in
  multi-antenna systems,'' \emph{IEEE Transactions on Wireless Communications},
  vol.~7, no.~2, pp. 714--725, 2008.

\bibitem{tulino2004random}
A.~M. Tulino and S.~Verd{\'u}, ``Random matrix theory and wireless
  communications,'' \emph{Foundamental Trends Communication Information
  Theory}, vol.~1, no.~1, pp. 1--182, Jun 2004.

\end{thebibliography}

\end{document}